\PassOptionsToPackage{dvipsnames}{xcolor}
\documentclass{aa}

\usepackage{graphicx}
\usepackage{txfonts}
\usepackage{orcidlink}
\usepackage{natbib}
\usepackage{placeins}
\usepackage[utf8]{inputenc}
\usepackage{hyperref}
\bibpunct{(}{)}{;}{a}{}{,} 

\newcommand{\ME}{M_{\oplus}}
\newcommand{\RE}{R_{\oplus}}
\newcommand{\RS}{R_{\odot}}
\newcommand{\msol}{R_{\odot}}
\newcommand{\ms}{\mathrm{m\,s^{-1}}}
\newcommand{\kms}{\mathrm{km\,s^{-1}}}
\newcommand{\gcc}{\mathrm{g\,cm^{-3}}}

\begin{document}

    \title{Revisiting TOI-4438 and TOI-442 planetary systems with new observations from SPIRou and TESS}
    \titlerunning{Revisiting TOI-4438 and TOI-442}
    
\author{
J. Serrano Bell\inst{1,2}\orcidlink{0000-0002-8397-557X}\and
G. H\'ebrard\inst{3,4}\and
E. Martioli\inst{5,3}\orcidlink{0000-0002-5084-168X}\and
R. F. Díaz\inst{1,2}\orcidlink{0000-0001-9289-5160}\and
L. de Almeida\inst{6}\orcidlink{0000-0001-8179-1147}\and
D. Lorenzo-Oliveira\inst{5}\and
A. Salmi\inst{7}\and
C. Dorn\inst{7}\orcidlink{0000-0001-6110-4610}\and
M. Valatsou\inst{7}\orcidlink{0009-0005-3682-1593}\and
A. Carmona\inst{10}\orcidlink{0000-0003-2471-1299}\and
M. Ould-Elhkim\inst{10}\orcidlink{0000-0003-2471-1299}\and
L. Arnold\inst{8}\and
\'E. Artigau\inst{9}\orcidlink{0000-0003-3506-5667}\and
I. Boisse\inst{10}\and
X. Bonfils\inst{11}\orcidlink{0000-0001-9003-8894}\and
C. Cadieux\inst{9,17}\orcidlink{0000-0001-9291-5555}\and
Z. Chakir\inst{9,13}\orcidlink{0009-0002-2901-3908}\and
N. J. Cook\inst{9}\orcidlink{0000-0003-4166-4121}\and
X. Delfosse\inst{11}\orcidlink{0000-0001-5099-7978}\and
J.-F. Donati\inst{12}\orcidlink{0000-0001-5541-2887}\and
R. Doyon\inst{9}\and
N. Heidari\inst{3}\and
J. M. Jenkins\inst{14}\orcidlink{0000-0002-4715-9460}\and
F. Kiefer\inst{15}\and
S. Lafrance\inst{9}\orcidlink{0009-0001-1040-7889}\and
A. L'Heureux\inst{9}\orcidlink{0009-0005-6135-6769}\and
C. Moutou\inst{12}\orcidlink{0000-0002-2842-3924}\and
J. Morneau\inst{9,16}\orcidlink{0009-0003-1186-5237}\and
X. Vandelac\inst{9,16}\orcidlink{0009-0003-5903-5224} 
}

\institute{
\inst{1}  Instituto de Ciencias Físicas (ICIFI-CONICET), ECyT-UNSAM, Campus Miguelete, 25 de Mayo y Francia, (1650) Buenos Aires, Argentina \\
\inst{2} Instituto Tecnológico de Buenos Aires (ITBA), Buenos Aires C1437, Argentina \\
\inst{3} Institut d'astrophysique de Paris, CNRS, UMR 7095, Sorbonne Universit\'{e}, 98 bis bd Arago, 75014 Paris, France. \\
\inst{4} Observatoire de Haute-Provence, St Michel l'Observatoire, France. \\
\inst{5} Laborat\'orio Nacional de Astrofisica, Rua Estados Unidos 154, 37504-364, Itajubá - MG, Brazil \\
\inst{6} SOAR Telescope/NSF’s NOIRLab, Avda Juan Cisternas 1500, 1700000, La Serena, Chile \\
\inst{7} Institute for Particle Physics and Astrophysics, ETH Zurich, CH-8093 Zurich, Switzerland \\
\inst{8} Canada France Hawaii Telescope (CFHT) Corporation, UAR2208 CNRS-INSU, 65-1238 Mamalahoa Hwy, Kamuela 96743 HI, USA \\
\inst{9} Trottier Institute for Research on Exoplanets and Department of Physics, Universit\'{e} de Montr\'{e}al, 1375 Ave Th\'{e}r\`{e}se-Lavoie-Roux, Montr\'{e}al, QC, H2V 0B3, Canada \\
\inst{10} Aix Marseille Univ, CNRS, CNES, LAM, Marseille, France \\
\inst{11} Univ. Grenoble Alpes, CNRS, IPAG, F-38000 Grenoble, France\\
\inst{12} Université de Toulouse, CNRS, IRAP, 14 avenue Belin, 31400 Toulouse, France \\
\inst{13} Department of Physics, McGill University, Ernest Rutherford Physics Building, 3600 University Street, Montr\'{e}al, QC, H3A 2T8, Canada \\
\inst{14} NASA Ames Research Center, Moffett Field, CA 94035, USA \\
\inst{15} LESIA, Observatoire de Paris, Université PSL, CNRS, Sorbonne Université, Université Paris Cité, 5 place Jules Janssen, 92195 Meudon, France\\
\inst{16} \'{E}cole Polytechnique de Montr\'{e}al, 2500 Chem. de Polytechnique, Montr\'{e}al, QC, H3T 0A3, Canada \\
\inst{17} Observatoire de Gen\`eve, D\'epartement d’Astronomie, Universit\'e de Gen\`eve, Chemin Pegasi 51, 1290 Versoix, Switzerland \\
}

\abstract{We present a comprehensive re-analysis of two star-planet systems: TOI-4438, an M3.5V star hosting a mini-Neptune in a 7.4-day orbit, and TOI-442, an M1V star with a 4-day period planet located within the hot Neptune desert. Both systems were originally identified as transiting planet candidates by the Transiting Exoplanet Survey Satellite (TESS) and subsequently validated through the radial velocity (RV) method. Our work incorporates new TESS transit data and high-resolution spectroscopy from the SPIRou near-infrared (nIR) spectropolarimeter. We detect a persistent and relatively strong Zeeman signature in TOI-442, while TOI-4438 exhibits weaker and intermittent magnetic activity, and we infer the stellar rotation periods of both stars from the variability of the longitudinal magnetic field. We jointly fit photometry and RV models for each system. For TOI-4438\,b we combine archival CARMENES data with 81 SPIRou observations and five TESS sectors. This yields a refined planetary mass of $M_{\rm p} = 4.11^{+0.40}_{-0.38}\,\ME$ and a radius of $R_{\rm p} = 2.40^{+0.09}_{-0.10}\,\RE$, consistent with the previous estimate within 1.1$\sigma$ while improving by 53\% the precision on the mass and 22\% on the radius. 
For TOI-442\,b, we add 29 SPIRou RV measurements to an extensive archival dataset, significantly extending the temporal baseline. Incorporating this with a new TESS sector, we tighten the constraints on the planetary mass to $M_{\rm p} = 28.38^{+0.77}_{-0.73}\,\ME$ and radius to $R_{\rm p} = 4.25^{+0.10}_{-0.08}\,\RE$, which agrees to the previous values within 1.5$\sigma$ and improves the precisions by 46\% and 67\% respectively. We find no clear signs of additional planets in the currently available RV data, although we detect a single-transit event in the TOI-4438 light curve. We compare various RV models and find that those accounting for stellar variability-induced signals yield improved constraints on the planetary parameters.
}
\maketitle
\section{Introduction}
M dwarfs are the most common type of star in the solar neighborhood \citep{Reyle2021}. Their small radii and low masses make them excellent targets for exoplanet detection because, for a planet of a given period, mass and radius they produce deeper transits and larger RV signals than solar-type hosts, facilitating the discovery of small, potentially habitable worlds. Radial velocity surveys have revealed a high occurrence rate of low-mass planets and temperate-zone companions around M dwarfs \citep[e.g.,][]{Mignon2025, Bonfils2013}. However, their higher magnetic activity and the fact that their emission spectrum peaks in the infrared, where there is significant telluric contamination, pose challenges to achieving the same RV precision as in solar-type stars. The Transiting Exoplanet Survey Satellite \citep[TESS;][]{Ricker2015} has expanded the sample of known exoplanets around M dwarfs ($\sim$176 TESS planets as of May 2026\footnote{NASA exoplanet archive: \url{https://exoplanetarchive.ipac.caltech.edu/} \citep{Christiansen_2025}}). With its mission extensions, many targets have been observed multiple times, allowing for better constraints on transit depths, planetary radii, and the identification of additional signals. These constraints are essential to situate M-dwarf planets within the broader exoplanet population, where key structures in parameter space remain poorly understood.

Sub-Neptunes and super-Earths, with sizes between Earth and Neptune, are incredibly common \citep[e.g.,][]{Howard2012, Fressin2013}, yet their nature remains elusive due to their absence in our own Solar System. These two types are separated by the radius valley \citep{Fulton2017}, a feature in the planet size distribution likely shaped by atmospheric loss \citep[e.g.,][]{VanEylen2018, Gupta2019} and core composition \citep[e.g.,][]{Venturini2020, Luque2022}, whose origin remains an active area of research. Another striking feature of the exoplanet population is the hot Neptune desert, a region in parameter space where close-in Neptune-sized planets are unexpectedly rare \citep{Mazeh2016, Owen2017, Owen2018}, with only a handful of planets known to reside within it \citep[e.g.,][]{West2019,Armstrong2020,Osborn2023,Nabbie2024,Hacker2024}. While the exoplanet sample will continue to expand with missions such as PLATO, Gaia, and Roman, progress on population origins requires more planets with well-constrained atmospheres. Facilities such as NASA's James Webb Space Telescope (JWST), the upcoming ESA's ARIEL mission and ESO's Extremely Large Telescope (ELT) or the planned Habitable Worlds Observatory (HWO) will enable this through transmission, emission, and reflected-light spectroscopy. 

In preparation for future atmospheric studies, the precise characterization of targets plays a key role. \citet{Changeat2020} explore how uncertainties in planetary mass affect atmospheric retrievals, highlighting that accurate mass estimates (ideally $\le$10\,\%) are essential, especially when clouds are present, to resolve key atmospheric properties such as mean molecular weight. \citet{Damiano2025} also find that a mass uncertainty $\le$10\% is needed to avoid biases in the determination of the dominant atmospheric species, although the study is focused on Earth-like planets. 

In this work we revisit two systems with new TESS photometry and RV observations from the SPIRou infrared spectropolarimeter. TOI-4438\,b was characterized by \citet{Goffo2024} as a volatile-rich mini-Neptune with a $7.45$-day period transiting a quiet, slow-rotating M3.5\,V star. Their analysis yielded a planetary radius of $2.52\pm0.13\,\RE$ and a mass of $5.4\pm1.1\,\ME$, corresponding to a bulk density of 1.85$^{+0.51}_{-0.44}\,\gcc$. These parameters suggested a low-density planet consistent with a volatile envelope. The planet was subsequently revalidated using multi-color transit photometry by \cite{PelaezTorres2024}, who measured a radius of $2.49\pm0.13\,\RE$.

TOI-442\,b was characterized by \citet{Dreizler2020} as a Neptune-sized planet with a $4.05$-day period transiting an early M dwarf. They reported a planetary mass of $30.8\pm1.5\,\ME$ and a radius of $4.7\pm0.3\,\RE$, resulting in a bulk density of $1.7\pm0.37\,\gcc$. These values indicated a low-density composition. It is presented as a favorable candidate for atmospheric studies given the host star brightness and planet size, but also because its location in the edge of the hot-Neptune desert \citep{Mazeh2016}.

\section{Observations}
\label{sec:observations}
\subsection{TESS photometry}
\label{subsec:tessphotometry}
TOI-4438 was observed by TESS between July 2021 and June 2022 in sectors 40, 52, and 53 at 2-min cadence, and later in June and July 2024 in sectors 79 and 80, when it was included among the 20-s cadence targets of the second extended mission. TOI-442 was observed in November 2018 (sector 5) and again in December 2020 (sector 32), both at 2-min cadence. In this work we analyze all available sectors for both targets (TOI-4438: 40, 52, 53, 79, 80; TOI-442: 5, 32), while the sectors not analyzed in the previous literature are sectors 79 and 80 for TOI-4438 and sector 32 for TOI-442.

\begin{figure*}[h!]
\centering
\includegraphics[width=\hsize]{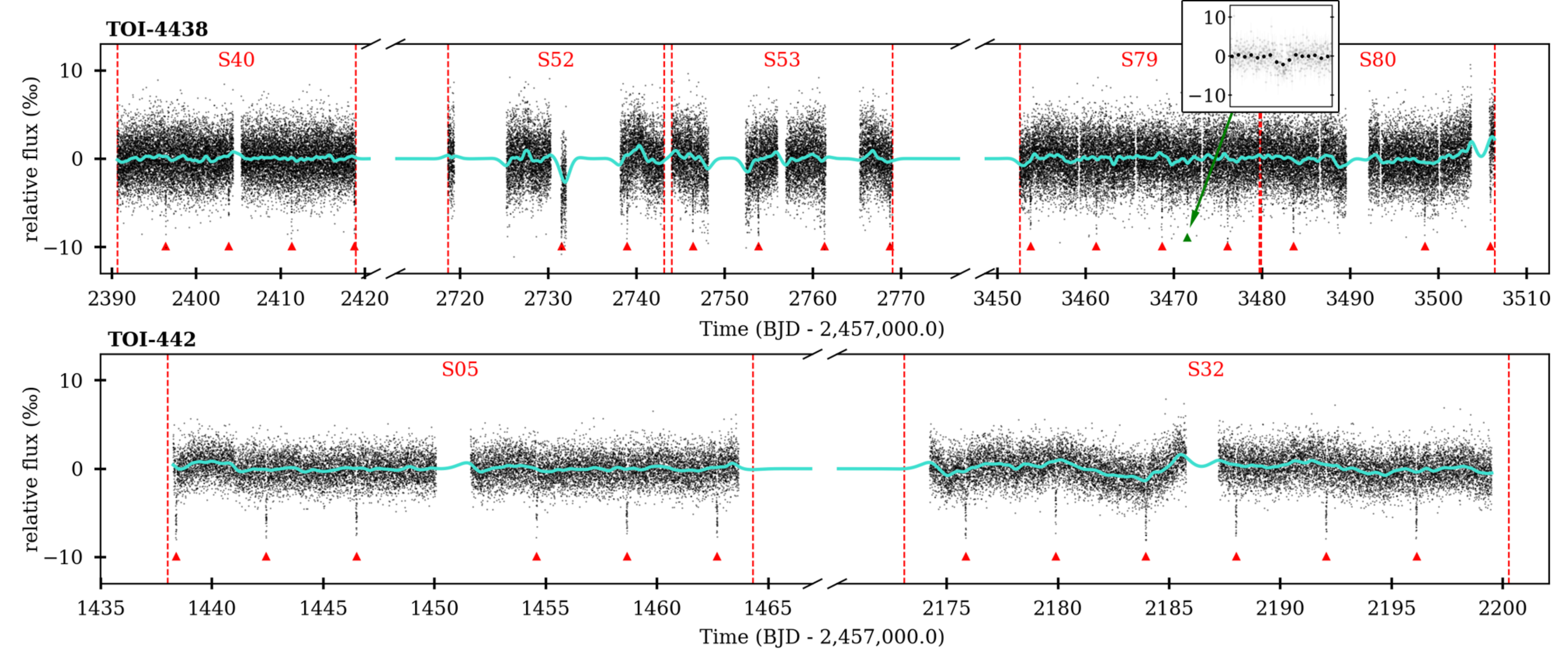}
\caption{TESS light curves. \textit{Upper panel:} PDCSAP flux for TOI-4438 with 2-min cadence. \textit{Lower panel:} PDCSAP flux for TOI-442 with 2-min cadence. The triangular markers indicate transits (see Sect.~\ref{subsec:lc_analysis}), in red for TOI-4438\,b and TOI-442\,b, and in green for the single transit (see Sect.~\ref{subsec:lc_analysis}) observed in TOI-4438 (also shown in the zoomed-in view including both the raw and binned light-curve data). The GP model (Sect.~\ref{subsec:lc_analysis}) is shown in cyan.}
\label{fig:tess_sectors}
\end{figure*}

The TESS photometric data were processed by the NASA Ames Science Processing Operations Center \citep[SPOC;][]{Jenkins2002,Jenkins2010, Jenkins2016, Jenkins2020}. We downloaded the light curves produced by the SPOC pipeline from the Mikulski Archive for Space Telescopes (MAST), in particular the flux from the Presearch Data Conditioning Simple Aperture Photometry (PDCSAP), which is corrected for crowding contamination, instrumental trends, and noise \citep{Stumpe2012, Stumpe2014, Smith2012}. We normalize the flux and remove all NaNs and 5$\sigma$ outliers using the \texttt{Lightkurve} package \citep{Lightkurve2018}. The standard deviation of the out-of-transit flux is 2.3\,ppt for TOI-4438 and 1.6\,ppt for TOI-442. Figure~\ref{fig:tess_sectors} shows the resulting PDCSAP light curves. The gaps in sectors 52 and 53 come from strong systematics that were removed.

Given the large pixel size of 21" of the TESS cameras, there is always the risk of flux contamination by close sources present in the Simple Aperture Photometry (SAP), the PDCSAP flux is already corrected for crowding, and the contamination ratios reported are 1.9\% for TOI-4438 and 0.016\% for TOI-442. The photometric mask used for the Simple Aperture Photometry (SAP) can be seen in the target pixel file images (Appendix \ref{apx:figs}, Fig. \ref{fig:tpf}) along with the positions of Gaia Data Release 3 \citep[DR3;][]{Gaia2023} sources. The difference image analysis provided in the multi-sector Data Validation reports for TOI-442 and TOI-4438 located the host stars within 1.95"\,$\pm$\,2.5", and 0.88"\,$\pm$\,2.7", respectively, of the transit source locations \citep{Twicken2018, Li2019}.

\subsection{SPIRou spectropolarimetry}
\label{subsec:spirou}
SPIRou\footnote{\url{http://spirou.irap.omp.eu} and \url{https://www.cfht.
hawaii.edu/Instruments/SPIRou/}} (SpectroPolarimètre InfraRouge) is a near-infrared spectropolarimeter and precision velocimeter installed at the 3.6-m Canada-France-Hawaii Telescope \citep{Donati2020}. Covering the YJHK bands (0.95–2.5~$\mu{\rm m}$) in a single exposure, it combines high-resolution velocimetry and spectropolarimetry. SPIRou is designed to detect and characterize planetary systems around nearby M dwarfs, and to study magnetic fields and star-planet formation processes. It achieves a resolving power of up to $70,000$ and a velocity precision down to $2\,\ms$ \citep{Donati2020, Cook2022}. The instrument supports extensive science programs, including the SPIRou Legacy Survey, which focuses on discovering planets, studying stellar magnetism, and characterizing exoplanet atmospheres. 

TOI-442 and TOI-4438 were observed under the large programs SPIRou Legacy Survey (SLS\footnote{\url{http://spirou.irap.omp.eu/Observations/
The-SPIRou-Legacy-Survey}}; PI: Jean-François Donati) and SPIRou Legacy Survey - Consolidation \& Enhancement (SPICE; PI: Jean-François Donati). TOI-4438 was observed from 2 June 2022 to 24 Sep. 2024, acquiring 83 spectra with signal-to-noise ratios (S/N) at 1745\,nm ranging from 130 to 220 and a median of 196. While TOI-442 was observed in two seasons separated by around five years, from 2 October 2019 to 11 December 2019 (14 epochs) and from 18 September 2024 to 22 December 2024 (25 epochs). The S/N for this spectra was between 120 and 210 with a median of 180.

SPIRou's effectiveness in characterizing systems including transiting planets was previously demonstrated in various studies, e.g. TOI-1759\,b \citep{Martioli2022}, TOI-3568\,b \citep{Martioli2024}, TOI-4860\,b \citep{Almenara2024}, TOI-1695\,b \citep{Kiefer2023}, or TOI-1452\,b \citep{Cadieux2022}.

\subsubsection{Radial velocities}
\label{subsec:spirou_rv}
The SPIRou spectra were reduced with APERO v0.65 \citep{Cook2022}, and RVs were derived with the line-by-line method \citep[LBL;][]{Artigau2022}, which combines per-line velocities with outlier rejection and is effective against stellar, telluric, and instrumental systematics. The LBL method is based on the formalism of \citealt{Bouchy2001}, which requires a noiseless spectral template of the observed star. In practice, either a high-S/N combined spectrum of the target or a standard star of similar spectral type is used. We used Gl 699 and GL 846 as templates for TOI-4438 and TOI-442, respectively. The RVs were then corrected for the barycentric Earth radial velocity (BERV), Fabry-Pérot drift, and instrumental zero point using the Gaussian process (GP) calibration of \citet{Cadieux2022}. To limit residual telluric contamination, we inspected the RVs as a function of $V_{\rm tot}=V_{\rm sys}-{\rm BERV}$ \citep{Carmona2025} and rejected the TOI-442 measurements with $|V_{\rm tot}|<10\,\kms$ (nine points), which reduced the dispersion from $23.8$ to $12.5\,\ms$. No TOI-4438 measurements were removed, as all of them had $|V_{\rm tot}|>10\,\kms$; their dispersion is $7.5\,\ms$. The median RV uncertainties are $3.1\,\ms$ for TOI-4438 and $4.0\,\ms$ for TOI-442.

A byproduct of the LBL method is the differential line width (${\rm dLW}$; first defined by \citealt{Zechmeister2018}). The ${\rm dLW}$ scales with the variation in the line profile full width at half maximum (FWHM) and can therefore be used as a stellar activity indicator \citep{Artigau2022}. More recently, \cite{Artigau2024} extended the LBL formalism to measure ${\rm dTemp}$, the disk-averaged differential change in stellar effective temperature relative to a reference template, which is now a standard product of the APERO pipeline. The ${\rm dTemp}$ is sensitive to line contrast changes induced by stellar activity, such as cool spots and hot plages, and thus constitutes an additional activity indicator derived directly from the spectra. The time series data derived from SPIRou spectropolarimetric observations used in this work are available in Appendix~\ref{apx:spirou_data}.

\subsubsection{Polarimetry}
\label{subsec:spirou_polarimetry}
The SPIRou observations of TOI-442 and TOI-4438 were obtained in circular polarization mode (Stokes~V). Each observation typically consists of a sequence of four exposures taken at different orientations of the two rotating Fresnel rhombs, providing one polarimetric measurement per visit. The polarimetry is computed following the method of \citet{Donati1997} and described in more detail by \citet{Martioli2020,Cook2022}. From the 83 visits on TOI-4438, two of them did not include the complete canonical sequence of four exposures. For those, we did not compute polarimetry.

We analyzed the SPIRou polarimetric spectra using the \texttt{spirou-polarimetry}\footnote{\url{https://github.com/edermartioli/spirou-polarimetry}} package \citep{Martioli2022}, which implements the same algorithms used in the APERO pipeline \citep{Cook2022}. The Stokes~I, Stokes~V, and null-polarization spectra have been analyzed using the least-squares deconvolution (LSD) methods of \citet{Donati1997}. The line mask used in our LSD analysis for both targets was constructed from the VALD database \citep{Piskunov1995} and a MARCS model atmosphere \citep{Gustafsson2008} with a surface gravity of $\log{g} = 5.0$~dex and effective temperatures of 4000~K and 3500~K for TOI-442 and TOI-4438, respectively. We selected all lines deeper than 3\% and with a positive Landé factor ($g_{\rm eff} > 0$), yielding a total of 2460 atomic lines for TOI-442 and 1336 atomic lines for TOI-4438. The resulting LSD Stokes~I and Stokes~V profiles at each observing epoch for TOI-4438 and TOI-442 are shown on Figures~\ref{fig:lsd_img2_4438} and~\ref{fig:lsd_img2_442}, respectively. 
\subsection{Archival radial velocities}
\label{subsec:rvs}
We include the radial velocity data for TOI-4438 from the Calar Alto high-Resolution search for M dwarfs with Exoearths with Near-infrared and optical Echelle Spectrographs (CARMENES) at the 3.5-m telescope in the Calar Alto Observatory, Almería, Spain \citep{Quirrenbach2014, Quirrenbach2018}. The CARMENES VIS spectrograph covers the wavelength range 520--960 nm, while the CARMENES NIR covers the near-infrared range 960--1710 nm, with resolutions of $\approx$94600 and $\approx$80400 respectively. The data were collected between 21 May 2022 and 2 May 2023, with 34 observations with both NIR and VIS instruments. We use the 32 CARMENES VIS RVs published by \citet{Goffo2024} (two were discarded due to los S/N), with an average S/N of $\sim$47 at 740\,nm. The CARMENES NIR data were not published by \citet{Goffo2024}, who deemed the NIR S/N insufficient 
for RV analysis; we therefore do not include NIR data from this instrument in our analysis. See further details on this dataset in \cite{Goffo2024}.

For TOI-442 there is an extensive archive of radial velocities that includes data from both optical and nIR spectrographs (see \citealt{Dreizler2020}). Observations with CARMENES VIS and NIR began in February 2019 and ended in November 2019, comprising a total of $33$ spectra. The VIS channel has a typical S/N of $\sim63$ at $746\,{\rm nm}$. The NIR dataset was excluded from our analysis due to its significant lower data quality: it exhibits an RV scatter of $22\,{\rm m\,s^{-1}}$ (compared to $12\,{\rm m\,s^{-1}}$ for the optical RVs) and a median uncertainty of $13\,{\rm m\,s^{-1}}$, more than twice that of any other dataset included in this work. Then 19 more spectra were obtained in September 2019 with ESPRESSO \citep{Pepe2010} with a S/N about 30 per pixel at $573\,{\rm nm}$. The HIgh Resolution Echelle Spectrograph \citep[HIRES;][]{Vogt1994} on the Keck-I telescope was used to acquire 14 spectra between August and November 2019. Six observations were also taken with the Carnegie Planet Finder Spectrograph \citep[PFS;][]{Crane2006, Crane2008, Crane2010} at the Magellan 2 (Clay) telescope at Las Campanas Observatory in Chile, with a typical S/N of 25. Finally, nine RVs were obtained over nine nights distributed in a 53\,d time span with the iSHELL spectrograph installed on the NASA Infrared Telescope Facility \citep{Rayner2016}. The S/N for this spectra varies from 107 to 171 at $2.2\,\mu{\rm m}$. 


\section{Data analysis}
\label{sec:analysis}
\subsection{Magnetic activity and stellar rotation} 
\label{subsec:stellaractivity}

For both stars, the Stokes~V profiles exhibit time-variable Zeeman signatures. To illustrate the variable magnetic activity of each target, Fig.~\ref{fig:lsd_img1} presents the mean LSD profiles of both systems, with the left panel showing the TOI-4438 profiles computed over six distinct time intervals (corresponding to observing seasons), as indicated in the panel titles, and the right panel displaying the TOI-442 profiles averaged over two separate seasons as well.
For TOI-442, the Zeeman signature is consistently detected, with an increased amplitude in 2024 compared to 2019. TOI-4438 magnetic activity is less visible, possibly due to self-cancellation, complex configuration, or lower activity. During some periods (e.g., April–June 2023 and July–September 2024), no Zeeman signature is detectable, even in the averaged profiles plotted in Figure~\ref{fig:lsd_img1}. However, during other periods (e.g., March–June 2024), a clear Zeeman signature is present. These results indicate that TOI-4438 does have magnetic active regions, but they appear less frequently than in TOI-442 or are less visible.

To investigate the magnetic activity in TOI-442 and TOI-4438, we computed the disk-integrated longitudinal magnetic field, $B_\ell$, from the SPIRou LSD profiles following \citet{Donati1997} and \citet{Martioli2020}. The values of $B_\ell$ are provided in Appendix~\ref{apx:spirou_data}. This quantity traces the net line-of-sight magnetic field over the visible stellar hemisphere and is expected to vary with rotational phase due to the non-uniform distribution of magnetic regions, enabling an estimate of the stellar rotation period if periodic variability is present.

We modeled the $B_\ell$ variability using a Gaussian process (GP) regression following \citet{Martioli2022} and \citet{Fouque2023}. We adopted a quasi-periodic kernel \citep{Angus2018} implemented with \texttt{george}\footnote{\url{https://george.readthedocs.io/}} \citep{georgecode} to describe rotation-modulated signals. The prior distributions of the hyperparameters are listed in Table~\ref{tab:activitygpfitparams}. We fixed the smoothing parameter to $\beta=0.5$ to allow deviations from strict periodicity while avoiding overfitting. The decay timescale was assigned a uniform prior between 10 and 1000 days, consistent with typical lifetimes of magnetic regions in M dwarfs, excluding shorter timescales to prevent overfitting.

\begin{figure}
    \centering
    \includegraphics[width=0.9\hsize]{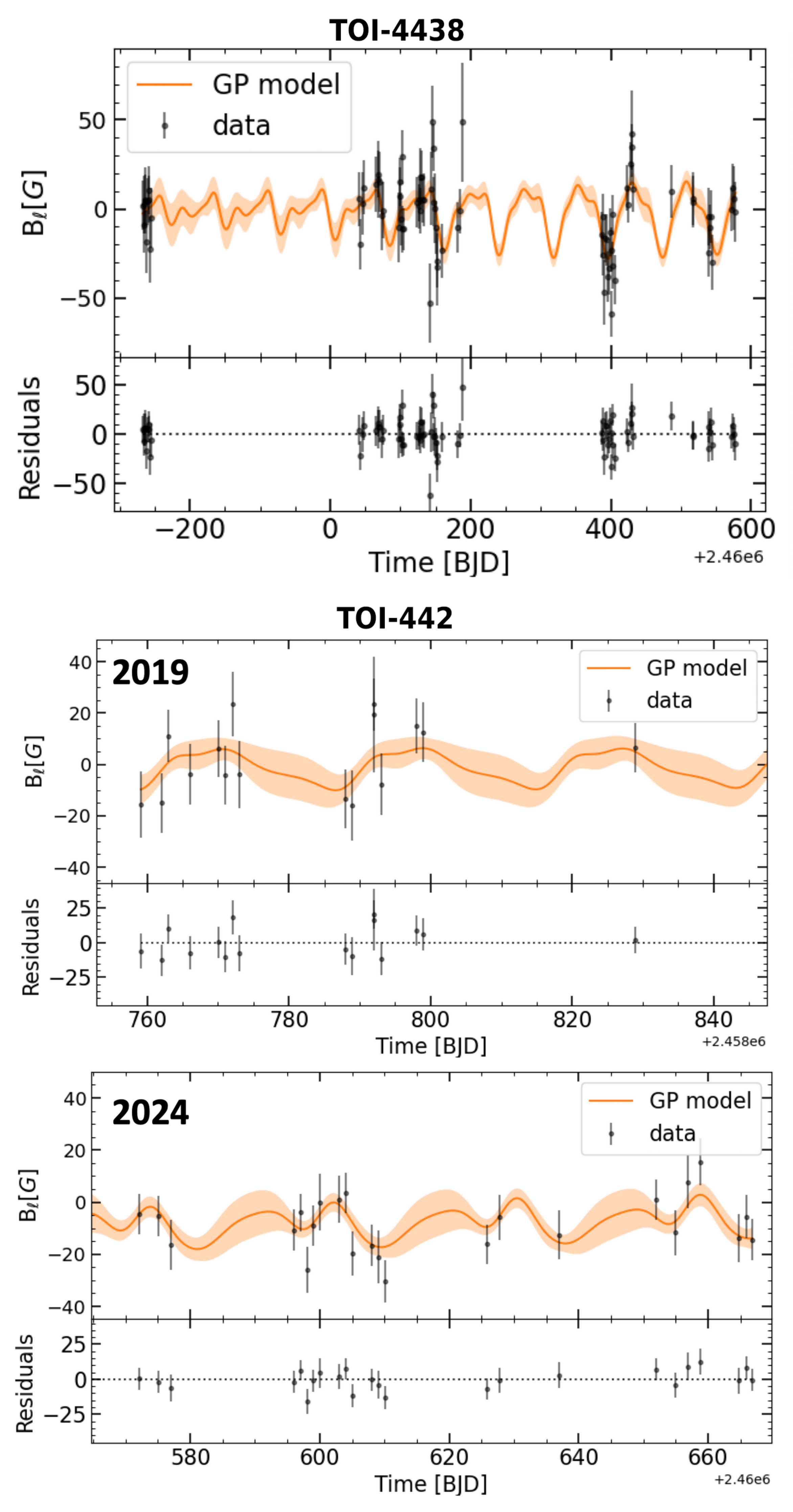}
    \caption{GP analysis of the time series of the longitudinal magnetic field ($B_\ell$). The black points represent the $B_\ell$ measurements derived from the LSD profiles of the SPIRou spectropolarimetric data, together with the best-fit GP model (orange lines).}
    \label{fig:bl_series}
\end{figure}

The posterior distributions of the GP hyperparameters were sampled using \texttt{emcee} \citep{foreman2013}, with 50 walkers, 5{,}000 burn-in steps, and 20{,}000 sampling steps. The posterior modes and $1\sigma$ uncertainties are reported in Table~\ref{tab:activitygpfitparams}, and the $B_\ell$ time series with GP fits are shown in Figure~\ref{fig:bl_series}.

We find distinct magnetic variability patterns for both stars, allowing us to constrain their rotation periods. TOI-4438 has $P_{\rm rot} = 77.6^{+0.9}_{-1.6}$\,d, a mean field of $B_\ell = -3 \pm 8$\,G, and an amplitude of $16^{+7}_{-4}$\,G. TOI-442 shows a faster rotation with $P_{\rm rot} = 28.6^{+1.6}_{-0.9}$\,d, a mean field of $-5 \pm 5$\,G, and an amplitude of $10^{+5}_{-3}$\,G. For both stars, the decay timescale is weakly constrained, with posterior modes of several hundred days. The rotation periods are consistent within $2\sigma$ with the previous estimates of $68\pm6$\,d and $33\pm3$\,d from photometric data \citep{Dreizler2020,Goffo2024}.

\subsection{Stellar parameters} 
\label{subsec:stellarparam}
Stellar parameters for TOI-442 and TOI-4438 were derived from a dedicated high-resolution spectroscopic analysis of the SPIRou data. Our procedure follows a three-step procedure: (i) global $\chi^2$ minimization over a pre-computed grid of synthetic spectra to locate the high-probability region; (ii) Markov Chain Monte Carlo (MCMC) sampling of the five grid parameters ($T_{\rm eff}$, $\log g$, [M/H], $\alpha$, $v_{\rm mic}$) using \texttt{emcee} \citep{emcee2013} to map posteriors and covariances; and (iii) a final non-linear least-squares refinement with MPFIT within iSpec \citep{ispec2019} to obtain the adopted values and formal uncertainties. Synthetic spectra were generated with MARCS \citep{marcs2008} atmospheres and MOOG\footnote{https://www.as.utexas.edu/~chris/moog.html} \citep{moog2012} radiative transfer; the line list is our internal list augmented by \cite{Sarmento2021}. We generated a grid of $\approx1.17\times10^{6}$ spectra, which provided the starting points for the MCMC runs and the MPFIT refinement. The derived parameters are presented in Table~\ref{tab:stellar_params}.

\begin{table*}
\centering
\footnotesize
\caption{Host stars parameters.}
\label{tab:stellar_params}
\begin{tabular}{lcccc}
\hline\hline
Parameter & TOI-4438 & Source & TOI-442 & Source \\
\hline
\multicolumn{5}{c}{Identifiers} \\
TIC & 22233480 & TESS & 70899085 & TESS  \\
2MASS & J18011609+3535514 & 2MASS & J04164560-1205023 & 2MASS  \\
Gaia DR3 & 4606416269652179328 & Gaia & 3189306030970782208 & Gaia  \\
\hline
\multicolumn{5}{c}{Magnitudes} \\
V (mag) & $13.69 \pm 0.05$ & TIC & $12.49 \pm 0.04$ & TIC  \\
G (mag) & $12.504 \pm 0.003$ & Gaia & $11.722 \pm 0.003$ & Gaia \\
J (mag) & $9.695 \pm 0.021$ & 2MASS & $9.493 \pm 0.024$ & 2MASS  \\
\hline
\multicolumn{5}{c}{Stellar parameters} \\
$\rm Sp. Type$  & M3.5V & \cite{Reid2003} & M1V & \cite{Gore2024} \\
$T_{\rm eff}$ (K) & $3501 \pm 70$ & This work & $4123 \pm 44$ & This work  \\
$\log g$ (cgs) & $4.67 \pm 0.11$ & This work & $4.60 \pm 0.08$ & This work \\
$[{\rm M/H}]$ (dex) & $0.07 \pm 0.10$ & This work & $0.05 \pm 0.10$ & This work \\
$\alpha$ (dex) & $-0.08 \pm 0.08$ & This work & $0.13 \pm 0.10$ & This work \\
$v_{\rm mic}$($\,\kms$)  & $2.35 \pm 0.42$ & This work & $1.05 \pm 0.34$ & This work \\
$L$ ($L_{\odot}$) & $0.01706 \pm 0.00007$ & \cite{Goffo2024} & $0.075 \pm 0.001$ & \cite{Dreizler2020} \\
$R$ ($R_{\odot}$) & $0.355 \pm 0.014$ & This work & $0.536 \pm 0.012$ & This work \\
$M$ ($M_{\odot}$) & $0.350 \pm 0.018$ & This work & $0.542 \pm 0.001$ & This work \\
$P_{\rm rot}$ (d) & $77.6_{-1.6}^{+0.9}$ & This work & $28.6_{-0.9}^{+1.6}$ & This work  \\
$\rm Age$ (Gyr) & $4.8 \pm 1.7$ & This work & $2.9 \pm 0.9$ & This work \\
\hline
\end{tabular}
\tablefoot{The age for both systems was estimated following the age-rotation empirical relationships for early type M dwarfs from \cite{Engle&Guinan2023}.}
\end{table*}

The solution converged to values consistent with published optical spectroscopy and spectral energy distribution (SED) constraints. The relatively large $v_{\rm mic}$ found for TOI-4438 is required by our line list and radiative-transfer setup to match observed line widths. Stellar radii were derived using the Stefan-Boltzmann law with the $T_{\rm eff}$ obtained in this work, and the bolometric luminosities reported by \cite{Goffo2024} and \cite{Dreizler2020}. Stellar masses for both TOI-442 and TOI-4438 were then computed from the empirical mass-radius relation for M dwarfs of \cite{Schweitzer2019}, yielding to $R_{\star} = 0.536 \pm 0.012 \, R_\odot$ and $M_{\star} = 0.542 \pm 0.017 \, M_{\odot}$ for TOI-442 and  $R_{\star} = 0.355 \pm 0.014 \, R_\odot$ and $M_{\star} = 0.350 \pm 0.018 \, M_{\odot}$ for TOI-4438. Independent estimates based on a mass-luminosity relation and on the spectroscopic $\log g$ are fully consistent with these values. 
\subsection{Periodogram analysis}
\label{subsec:periodogram}
We ran generalized Lomb-Scargle\footnote{We used the astropy implementation \citep{Astropy2022}.} \citep[GLS;][]{GLS2009} periodograms of the SPIRou RVs, ${\rm dLW}$, ${\rm dTEMP}$ and the full RV time series for both targets. 

\begin{figure}
\centering
\includegraphics[width=0.9\hsize]{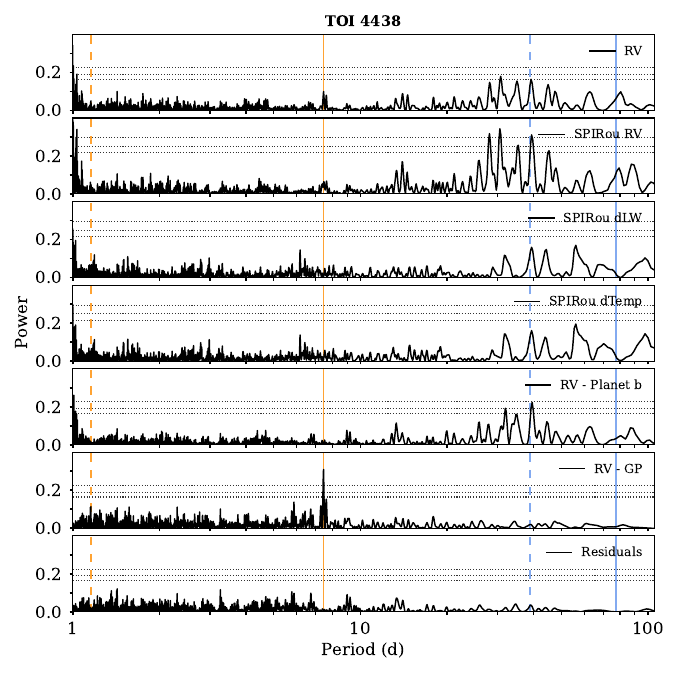}
\caption{GLS periodograms for TOI-4438 for (from top to bottom): combined RVs dataset, SPIRou RVs, SPIRou ${\rm dLW}$, SPIRou ${\rm dTEMP}$, residuals from the Keplerian model of planet b, residuals from the GP model for stellar activity and residuals from the full final model. Horizontal dashed lines show the 1\%, 5\% and 10\% false alarm probabilities. The orange solid and dashed line mark the period of the TOI-4438\,b and its 1-day alias. The blue solid and dashed lines mark the stellar rotation period from Section~\ref{subsec:stellarparam} and its first harmonic.}
\label{fig:rvgls_4438}
\end{figure}

Figure~\ref{fig:rvgls_4438} presents the periodogram analysis of the TOI-4438 data. The $\sim$7.4-day signal associated with the transiting planet is not significantly detected in either the SPIRou RVs or the full RV dataset. Instead, both datasets show excess power between 20 and 100 days; in the SPIRou RVs, three peaks exceed the 1\% false-alarm probability (FAP) threshold and are located near half the stellar rotation period. The SPIRou activity indicators ${\rm dLW}$ and ${\rm dTEMP}$ also show excess power at longer periods. In the full RV dataset, the planetary signal also appears with a FAP below 10\%, and the strongest peaks lie near the stellar rotation period and its first harmonic.

\begin{figure}
\centering
\includegraphics[width=0.9\hsize]{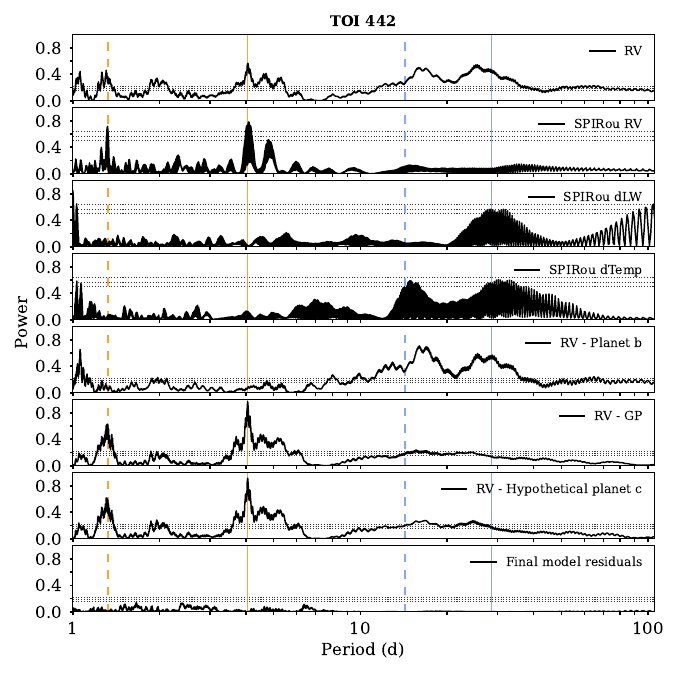}
\caption{GLS periodograms for TOI-442 for (from top to bottom): combined RVs dataset, SPIRou RVs, SPIRou ${\rm dLW}$, SPIRou ${\rm dTEMP}$, residuals from the Keplerian from planet b, residuals from the GP model for stellar activity, residuals from the Keplerian of the hypothetical planet c and residuals from the full final model. Horizontal dashed lines show the 1\%, 5\% and 10\% false alarm probabilities. The orange solid and dashed line mark the period of the TOI-442\,b and its 1-day alias. The blue solid and dashed lines mark the stellar rotation period from Section~\ref{subsec:stellarparam} and its first harmonic.}
\label{fig:rvgls_442}
\end{figure}

The periodograms for TOI-442 are shown in Figure~\ref{fig:rvgls_442}. In the SPIRou data, the signal of TOI-442\,b ($\sim$4.05\,days) is detected with a FAP$<$1\% in the RVs. The ${\rm dLW}$ shows a clear peak near the stellar rotation period, and ${\rm dTEMP}$ shows significant peaks both at the stellar rotation period and its second harmonic. The analysis of the full RV time series reveals a strong signal at the expected period of the transiting planet. Other significant peaks are present in this series: at $\sim$1.3\,d which can be associated with a 1-day alias of the planet's $\sim$4-day period, at half the period of the planet $\sim$2\,d, and then there are broad peaks around $\sim$16 and $\sim$26\,days, that could be related to stellar activity or variability as they are near the rotation period and its first harmonic. Given the large time gap between the second SPIRou season and the remaining RVs, we also computed a periodogram restricted to the pre-gap data, but found no reduction in the aliasing structure seen in the periodogram.

\subsection{Light curves} 
\label{subsec:lc_analysis}
We independently searched the TESS light curves of TOI-4438 and TOI-442 (Sect.~\ref{subsec:tessphotometry}) to search for transit signals using the Transit Least Squares algorithm \citep[TLS;][]{TLS2019}. For TOI-4438, we scanned periods from 0.6 d to half the time-series baseline with 64 trial durations and recovered TOI-4438\,b with a signal detection efficiency (SDE) of 59.2 (\citealt{Kovacs2002}), corresponding to 17 transits, including seven not analyzed previously. After masking these transits and re-running the search, no additional periodic signal was found. We nevertheless identify a single transit-like event in sector 79 (see Fig.~\ref{fig:tess_sectors}); a box least-squares \citep[BLS;][]{Kovacs2002} fit gives a depth of 0.21$\pm$0.03\%, a duration of 2.2\,h, and a mid-transit time of 3471.559 TESS barycentric Julian date (TBJD). Checks of sector release notes and quality flags show no indication that this feature is instrumental. In the case of TOI-442, we search for periods between 0.6 days and half the time span of the TESS light curve with a grid of 43 durations. We strongly detect the transit signal of TOI-442\,b with a SDE of 64.4 and a total of 12 transits visible in the two sectors (five of which were not analyzed in the literature). Again, we remove the in-transit points and run another search for possible smaller shallower signals, but do not find anything significant. The detected transits can be seen in Fig.~\ref{fig:tess_sectors}.

Although PDCSAP light curves are corrected for instrumental trends, some low-amplitude variability persists, this can be attributed to residual instrumental noise or stellar variability. To investigate whether the residual variability is modulated by stellar rotation, we performed a GLS periodogram analysis of the out-of-transit data for both targets. No significant periodicity was found, indicating that the variability is aperiodic and consistent with residual instrumental noise, granulation or low-level stellar surface activity. 

In order to improve the transit depth fit in the joint model of Sect. \ref{subsec:joint}, we model and correct this variability using a GP. We use the stochastically driven damped simple harmonic oscillator (SHO) kernel \citep{Foreman2017} implemented in \texttt{celerite}, which has a power spectral density (PSD) given by:
\begin{equation}
     S(\omega) = \sqrt{\frac{2}{\pi}} \frac{S_0 \,\omega_0^4}{(\omega^2-\omega_0^2)^2+\omega_0^2 \omega^2/ Q^2}
\end{equation}
where $\omega_0$ is the frequency of the undamped oscillator, $S_0$ is proportional to the power at $\omega_0$ and Q is the quality factor of the oscillator. A special case results when the quality factor is chosen to be $Q$ = 1/$\sqrt{2}$, the resulting kernel is well suited for the modeling of stellar variability in light curves as its PSD matches that of stellar granulation \citep{Foreman2017,Barros2020}, in that particular case the PSD simplifies to:
\begin{equation}
    \rm S(\omega) = \sqrt{\frac{2}{\pi}} \frac{S_0}{(\omega/\omega_0)^4+1}
\end{equation}
and the kernel in this case is given by:
\begin{equation}
    k(\tau) = \rm S_0 \, \omega_0 \, e^{-\frac{1}{\sqrt{2}}\omega_0\tau} \cos{ \left( \frac{\omega_0\tau}{\sqrt{2}}-\frac{\pi}{4} \right) }
\end{equation}
where $\tau$ is the time lag between two points. Since $Q$ is fixed, the remaining parameters of the kernel are the standard deviation of the process $\sigma_{\rm GP} = \sqrt{S_0 \omega_0 Q}$ and the undamped period of the oscillator $\rho_{\rm GP}$ = 2$\pi/\omega_0$. We also include a jitter term to the diagonal of the covariance matrix added quadratically to the flux uncertainty of each point. We fit this model to the light curves of both targets with the data within transits removed and fit it by finding the local maximum a posteriori (MAP). Then we compute the predictions for the on-transit points, the resulting model is shown overplotted to the light curves on Fig.~\ref{fig:tess_sectors} and the priors and MAP values in Table~\ref{table:gp_tess}. Finally, the GP model is subtracted from the PDCSAP flux. The recovered timescales of $\sim12\,\rm{h}$ are significantly longer than expected for granulation in M dwarfs \citep{Beeck2013}, and are more likely associated with residual instrumental systematics or other uncharacterized correlated noise in these timescales.

\begin{table}
\caption{Parameters of the GP model for TESS light curves.}           
\label{table:gp_tess}      
\centering         
\footnotesize                             
\begin{tabular}{l c c c}          
\hline\hline                        
Parameter & Prior & \multicolumn{2}{c}{MAP value} \\    
\hline               
     &  & TOI-4438 & TOI-442  \\  
    $\sigma_{\rm GP}$\,(ppt) & $\mathcal{U}(0, 10)$ & 0.54 & 0.47  \\   
    $\rho_{\rm GP}$\,(d) & $\mathcal{U}(0, 10)$ & 2.46 & 2.05 \\
    $Q$ & fixed & 1/$\sqrt{2}$ & 1/$\sqrt{2}$    \\
    timescale\,(d) & derived & 0.55 & 0.46 \\
    jitter\,(ppt) & $\mathcal{U}(0, 10)$ & 2.24 & 1.49 \\
\hline                                  
\end{tabular}
\tablefoot{$\mathcal{U}(a,b)$ represents a uniform distribution between $a$ and $b$. The timescale of the process is 2$Q/\omega_0$.}
\end{table}

\subsection{Joint transit and RV modeling}
\label{subsec:joint}
We infer system orbital and physical parameters with a Bayesian joint fit to the TESS light curves and RVs (Sect.~\ref{sec:observations}), sampled with MCMC. We implement the model in \texttt{PyMC\,5+}\footnote{\url{https://www.pymc.io/welcome.html}}, which includes a Hamiltonian Monte Carlo with a No-U-Turn Sampler \citep{Hoffman2011, Pymc2016}. The latter adaptively determines the integration path length, avoiding the need for manual tuning and improving the efficiency of the exploration of the posterior distribution, particularly in high-dimensional parameter spaces. We use the \texttt{exoplanet} \citep{Foreman2021} toolkit to build the Keplerian transit+RV model. 

A similar modeling was applied for both targets, unless specified otherwise. The photometry used is the GP-detrended TESS light curves from Sect. \ref{subsec:lc_analysis} keeping only one day around each transit, and is modeled using the time of mid transit $T_0$, orbital period $P$, impact parameter $b$, planet-to-star ratio $R_{\rm p}/R_\star$ and the quadratic limb-darkening coefficients $(q_1, q_2)$ following the parametrization from \citet{Kipping2013}. A flux offset is included and a TESS jitter term for white noise which is introduced in the model by adding it in quadrature to the flux uncertainties. The radius and mass of the host star are also fitted, with Gaussian priors informed by the analysis from Section \ref{subsec:stellarparam}. The orbital eccentricity $e$ and argument of periastron $\omega$, which enter both the transit and RV components of the model, are treated as free parameters. Although earlier analyses do not indicate significant eccentricity, we retain the eccentric parameterization in order to place upper limits on the eccentricities from the posterior distributions. We verified that circular-orbit solutions yield planetary parameters consistent within uncertainties with the eccentric ones. We adopt uniform priors on $e$ and $\omega$ between $[0, 0.5]$ and $[-\pi,\pi]$, respectively. For the RV part of the model we also fit the radial velocity semi-amplitude $K$, and include a per-instrument RV offset and jitter term. Because the SPIRou data extend both RV baselines by several years (Sect.~\ref{subsec:spirou}), we test models with and without a quadratic time term to capture possible long-term variability and compare their statistical support.

Correlated noise from stellar variability is common in M dwarfs and is often modeled using Gaussian processes (GPs) \citep[e.g.,][]{Martioli2022, Cadieux2022, Kiefer2023}. A quasi-periodic (QP) kernel is widely adopted to capture rotation-driven variability such as star spots. \citet{Stock2023} showed that QP-GP modeling yields tighter planetary parameters than purely Keplerian models and recovers rotation periods consistent with stellar values.  We include a GP component in the RV models using two kernels: (i) the SHO kernel described in Section~\ref{subsec:stellaractivity}, parameterized by the rotation period $P_{\rm rot}$ (with $P_{\rm rot} = 2\pi/\omega_0$), the damping timescale $\tau$, and the amplitude $\sigma_{\rm GP}$. And (ii), a double-SHO (dSHO) kernel implemented in \texttt{celerite}, combining two SHO terms with the second at the first harmonic of the first. This kernel has four free parameters: $\sigma_{\rm GP}$, $P_{\rm rot}$, the quality factor $Q_0$ (which is the quality factor of the secondary oscillation minus $1/2$)\footnote{\url{https://celerite2.readthedocs.io}}, the difference in quality factors $\Delta Q$, and the fractional amplitude $f$ of the secondary mode. This choice is motivated by the periodicity found near half the stellar rotation period in the RVs and activity indicators (Figs.~\ref{fig:rvgls_4438} and \ref{fig:rvgls_442}), and by its demonstrated performance in modeling stellar signals in both multi-dimensional and one-dimensional GP frameworks \citep{Bryant2022, Basant2025}.

Because activity-induced RV signals are inherently chromatic, we group the RVs into optical and near-IR datasets with independent GP components. The contrast of spots and plages varies with wavelength, and magnetic effects such as Zeeman broadening further introduce wavelength-dependent RV shifts \citep{Reiners2013, Larue2025}. However, the strength of the chromaticity depends on the details of the spots \citep{Reiners2010} and is not universally found in early to mid M dwarfs, especially if the RV variability is of low amplitude \citep[e.g.,][]{Cortes-Zuleta2023}. In our case, the RV scatter in nIR and optical bands suggests that chromaticity cannot be ruled out a priori: for TOI-4438, the CARMENES VIS scatter is $7.5$\,m\,s$^{-1}$ and $3.1$\,m\,s$^{-1}$ for SPIRou; while for TOI-442, it ranges between $12$ and $15$\,m\,s$^{-1}$ with some variation between instruments. We therefore adopt independent GP components for the optical and nIR datasets as a conservative choice, while acknowledging that the two components may converge to similar solutions if the activity signal is in practice achromatic. The nIR and optical GP components share the same hyperparameter $P_{\rm rot}$, which has a Gaussian prior informed by the posterior of the stellar rotation period obtained from the $B_\ell$ variability modeling in Section~\ref{subsec:stellaractivity}; all other parameters are independent for each band and have wide priors.

For each system, we test three noise models (K, K+SHO, K+dSHO), each with and without a quadratic RV drift term (D), yielding six models: K, K+D, K+SHO, K+SHO+D, K+dSHO, and K+dSHO+D. We sample posteriors with two MCMC chains of 6000 draws each (first 3000 for tuning), initialized at the MAP solution, and assess convergence using the Gelman-Rubin statistic and effective sample size.

\subsection{Model comparison and results}
\label{subsec:model_comparison}
To compare competing RV models (the light-curve model is identical in all cases), we use leave-one-out cross-validation (LOO-CV), a robust Bayesian criterion for model evaluation \citep{Vehtari2016, Gelman2013}. Unlike the Akaike information criterion or Bayesian information criterion, LOO-CV uses the full posterior and is well suited to models with different numbers of planets and noise treatments. It estimates the expected log predictive density (elpd) by iteratively predicting each held-out observation from fits to the remaining data, thereby penalizing over-complex models through predictive performance. Exact LOO-CV is rarely used because of the computational cost of refitting a model N times (for a dataset of N observations), however, there are workarounds. One such method uses Pareto-smoothed importance sampling (PSIS) to approximate elpd from posterior samples \citep{Vehtari2016}. We use the \texttt{Arviz} implementation of PSIS-LOO-CV\footnote{\url{https://python.arviz.org/en/stable/api/generated/arviz.loo.html}}\footnote{PSIS-LOO requires element-wise log-likelihoods, which are unavailable for GP models in the \textit{marginal} formulation. For those cases we compute LOO-CV analytically from the GP covariance matrix (Section 5.4.2 of \citealt{Rasmussen2006}).}. Models are then ranked by their LOO-CV scores. 

\subsubsection{Results for TOI-4438}
\label{subsec:toi4438_results}
For TOI-4438, we find that models incorporating stellar variability yield superior predictive accuracy (Fig.~\ref{fig:loocv}). Among these, the model with the single SHO kernel and drift achieved the highest score. Figure~\ref{fig:rvgls_4438} (panel 6) shows that subtracting the SHO-GP component removes the spurious long-period peaks and significantly enhances the planetary peak at the known 7.4-day period. Although the difference in LOO-CV scores of the SHO and dSHO kernels remain within the estimated uncertainties of the metrics, we adopt the \textbf{K+D+SHO} model as our final choice for TOI-4438.

\begin{figure*}[h!]
\centering
\includegraphics[width=0.8\hsize]{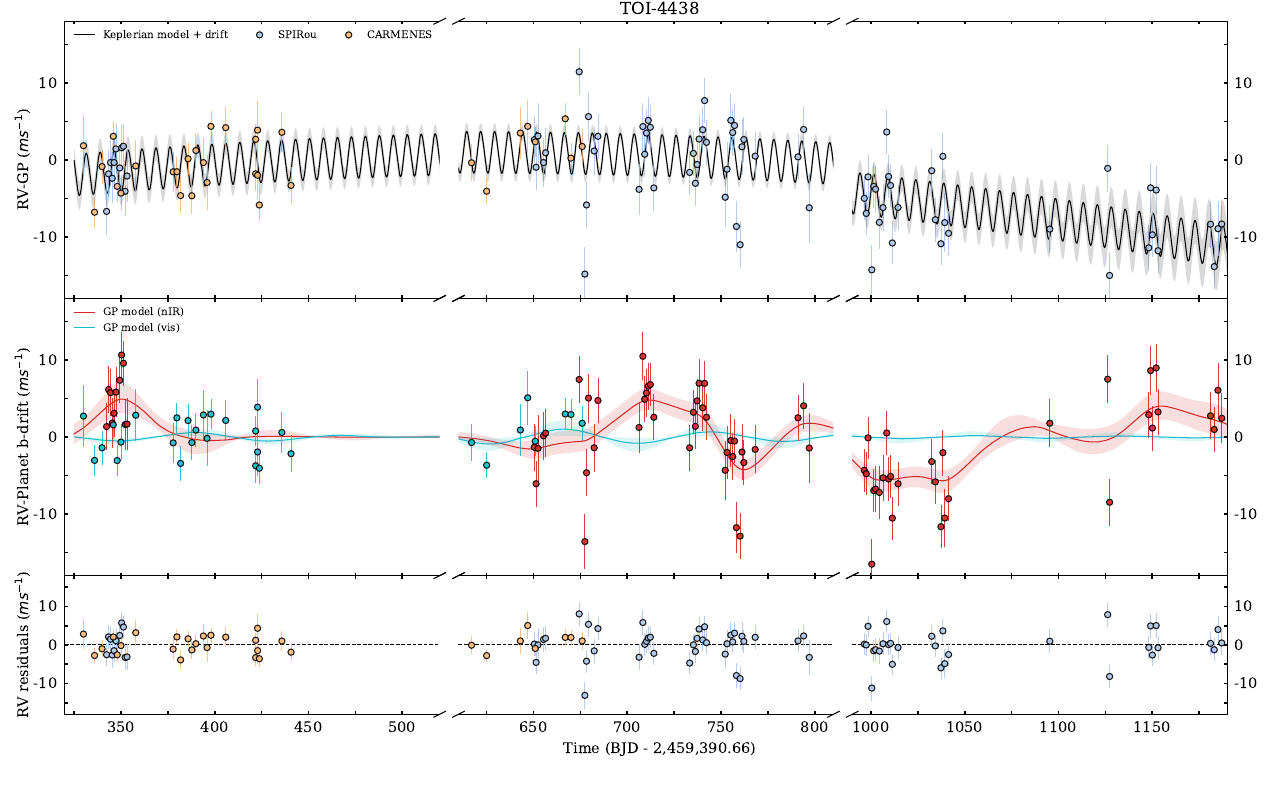}
\caption{Best-fit model for TOI-4438\,b, the solid black line and its gray overlay correspond to the median and 16th-84th percentile regions of the posterior. Top panel: Keplerian model and drift. Middle panel: SHO-GP model. Bottom panel: Full model residuals.}
\label{fig:rvfit_4438}
\end{figure*}

This model selection does not significantly affect the inferred planetary parameters, which remain consistent within uncertainties across all tested models. The main effect of including the GP is an improved treatment of stellar RV variability, leading to substantially lower RV jitter estimates. Figure~\ref{fig:rvfit_4438} presents the full RV model, while Figs.~\ref{fig:rvfit_phase} and \ref{fig:lcfit} show the phase-folded RV and transit fits. The full list of fitted and derived parameters is provided in Table~\ref{tab:results}.

\subsubsection{Results for TOI-442}
\label{subsec:toi442_results}
For TOI-442, the evidence strongly favors the more complex models: (i) All models including a correlated noise component achieve significantly better predictive accuracy (Fig.~\ref{fig:loocv}). The dSHO component captures signals near the rotation period and its first harmonic; removing it enhances the power of the transiting planet signal (see panel 6 of Fig.~\ref{fig:rvgls_442}). The posterior of $P_{\mathrm{GP}}$ (Fig.~\ref{fig:corner_442}) converges to the rotation period measured from the polarimetric data. (ii) The choice of noise model affects the precision of the planetary parameters. While the parameter values across different models remain statistically consistent, the uncertainties improve markedly when a GP is included. For example, the uncertainty in $M_{\rm p}$ decreases from 4.6\% (Keplerian-only) to 2.9\% with the SHO component, and to 2.7\% in the model with the dSHO GP. (iii) Instrumental jitter is substantially overestimated when correlated noise is ignored. In particular, the Keplerian-only model yields a jitter of $5.9\pm1$\,m\,s$^{-1}$ for ESPRESSO, whereas including a GP reduces this to a more realistic $1.0_{-0.4}^{+0.3}$\,m\,s$^{-1}$, consistent with previous studies \citep[e.g.][]{Pepe2021}.
Before selecting the final model for TOI-442, we first discuss the second-planet hypothesis in the following section.

\subsection{Search for additional planets}
\label{subsec:search}

\subsubsection{TOI-4438: single-transit candidate}
\label{subsec:toi4438_candidate}

In TOI-4438, subtracting the Keplerian signal of planet~b leaves two residual RV peaks with false-alarm probabilities of 5\%--10\%, both near the first harmonic of the stellar rotation period. These are captured by the SHO-GP component, and no significant peaks remain after removing the Keplerian+GP signals (Fig.~\ref{fig:rvgls_4438}, panels 5--7). We nevertheless detect the single transit-like event discussed in Section~\ref{subsec:lc_analysis} at 3471.559~TBJD, with depth $\sim2\,{\rm ppt}$, consistent with a $\sim1.8\,\RE$ planet. The target will be observed during TESS Cycle 9 during sector 118\footnote{According to the TESS point web tool: \url{https://heasarc.gsfc.nasa.gov/wsgi-scripts/TESS/TESS-point_Web_Tool/TESS-point_Web_Tool/wtv_v2.0.py/}}.

Although a single transit cannot validate a planet, it can constrain $P$ when stellar properties are known \citep{Seager2003}. Using the transit-geometry relation between duration, stellar density, and period (Eq.~3.4 of \citealt{Haswell2010}), and requiring no additional transits within the observed TESS baseline, we modeled the event in \texttt{exoplanet} assuming a circular orbit. We adopted Gaussian priors on stellar mass and radius from the joint fit of planet~b; sampled $\rho_\star$, $P$, $t_0$, $R_p/R_\star$, and $b$; fixed limb darkening to the joint-fit medians; imposed $P>32.7\,{\rm d}$ by rejecting trial periods that predict extra transits in the observed timestamps; and ran two MCMC chains of 2500 draws. We obtain $R_{\rm p}=1.8\pm0.1\,\RE$ and a long-tailed period posterior with median $P=74^{+152}_{-34}\,{\rm d}$. The fit is shown in Fig.~\ref{fig:single_transit}, and priors/posteriors in Table~\ref{tab:single_transit}. Using the empirical mass-radius relations from \citet{Muller2024}, we estimate $M_{\rm p}\approx5.8\pm1.2\,\ME$. Accounting for the uncertainties in both planetary mass and orbital period, the expected radial velocity semi-amplitude is $1 \lesssim K \lesssim 2.6$~m~s$^{-1}$, which is comparable to or smaller than the mean RV uncertainties of the data analyzed in this work (3.1 and 2.5~m~s$^{-1}$). Radial-velocity confirmation of this candidate is therefore challenging with the current data and would benefit from follow-up observations with higher-precision spectrographs.

\subsubsection{TOI-442: testing the second-planet hypothesis}
\label{subsec:toi442_planetc}

After subtracting a single Keplerian signal from TOI-442\,b, the RV-residual periodogram still shows excess power between 10 and 40 d, peaking near 16 d (Fig.~\ref{fig:rvgls_442}, panel 5). Because this period is close to the first harmonic of stellar rotation and also appears in ${\rm dTEMP}$, a stellar-activity origin is plausible, but a second planet remains possible. \citet{Dreizler2020} examined the same scenario and found that a two-planet fit would imply a Neptune-mass companion near a 4:1 resonance with planet~b; however, model comparison was inconclusive and they preferred the stellar-noise interpretation pending additional data. With the larger photometric and RV datasets now available, we revisit this possibility for TOI-442 by doubling the tested models (from six to twelve) to include a second Keplerian component, parameterized by $P_{\rm c}$, the mean anomaly at a reference epoch $\lambda_{0 \rm c}$, $K_{\rm c}$, $e_{\rm c}$, and $\omega_{\rm c}$. We adopt wide uniform priors for all parameters except $P_{\rm c}$, which we restrict to 14--18\,d, where the GLS residuals show the strongest excess power after removing planet~b (panel~5 of Fig.~\ref{fig:rvgls_442}). The broad 10--40\,d excess, rather than a sharp peak, renders the posterior of $P_{\rm c}$ sensitive to this prior choice.

\begin{figure}
\centering
\includegraphics[width=0.8\hsize]{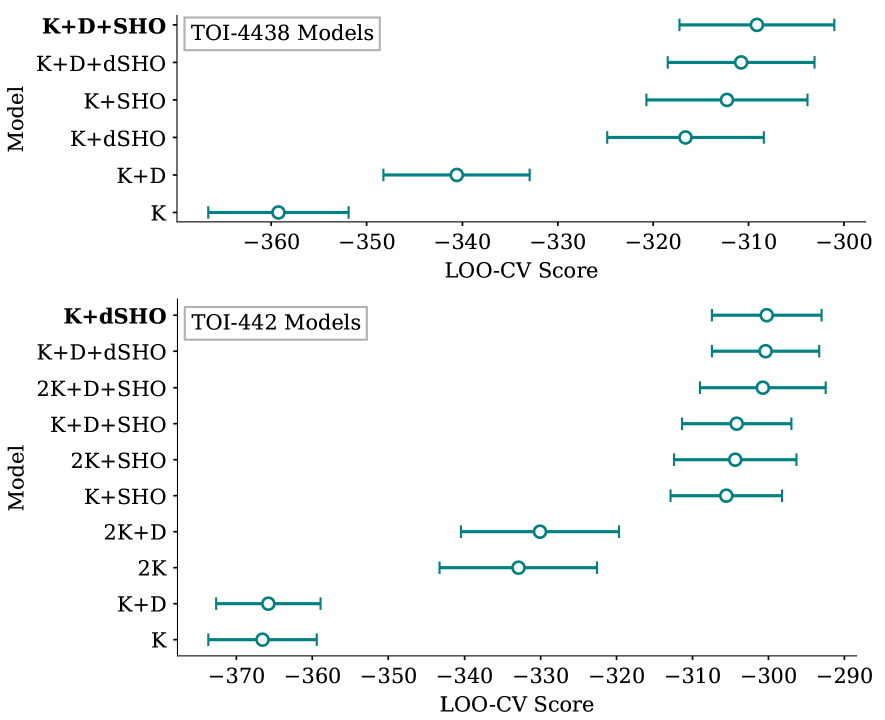}
\caption{Leave-one-out cross-validation (LOO-CV) scores with their estimated uncertainties for all evaluated models. For non-GP models, the scores are computed using the PSIS-LOO-CV implementation in \texttt{Arviz}; otherwise, the analytical LOO-CV expression is used (Section 5.4.2 from \citet{Rasmussen2006}).}
\label{fig:loocv}
\end{figure}

Among the models with a second Keplerian term (2K, 2K+D, 2K+SHO, and 2K+D+SHO), those including a SHO-GP are preferred by LOO-CV (Fig.~\ref{fig:loocv}). Two-planet models with dSHO are excluded because they do not converge, likely because the dSHO harmonic overlaps with the signal fitted by the second Keplerian. We note that the considerations (i), (ii), and (iv) from Sect.~\ref{subsec:toi442_results} also apply here. We find that including a drift does not materially change the hypothetical second-planet parameters but is slightly favored by the metric. Results for the two-planet models with drift are summarized in Table~\ref{table:planetc}, and the best-fit candidate-planet signal is shown in Fig.~\ref{fig:planetc_fit}.

A coplanar planet with this minimum mass should produce a detectable TESS transit depth ($\sim3.5\,{\rm ppt}$)\footnote{Using the mass-radius relations by \citet{Muller2024}.}, but no such event is seen; given the candidate period, at least four transits are expected within the TESS coverage. A mutual inclination could still suppress transits. In the SHO-GP solution, the period ratio with planet b is within 1.3\% of a 4:1 resonance. \citet{Dreizler2020} found a similar solution ($P_{\rm c}=16.06^{+0.10}_{-0.09}$\,d and $e_{\rm c}=0.12^{+0.16}_{-0.09}$) and showed that this commensurability is stable up to $e\sim0.4$. Such a third-order resonance can produce transit timing variations (TTVs) for non-zero eccentricity \citep{Agol2018}. For our 68\% highest density interval (HDI) on $e_{\rm c}$ (0.18--0.36), N-body integrations with the \texttt{REBOUND} package \citep{Rebound2012} predict TTV amplitudes of planet~b of $\sim$3--40\,min over the $\sim$800-day \textit{TESS} baseline.

To test this, we fit individual mid-transit times and obtain observed-minus-calculated (O--C) values of only a few minutes, consistent with zero at 1--2$\sigma$ (Fig.~\ref{fig:ttvs}, Appendix~\ref{apx:figs}). A Shapiro-Wilk test on the 12 timing residuals gives $p=0.46$, so we do not reject normality around zero.

\begin{figure*}[h!]
\centering
\includegraphics[width=0.8\hsize]{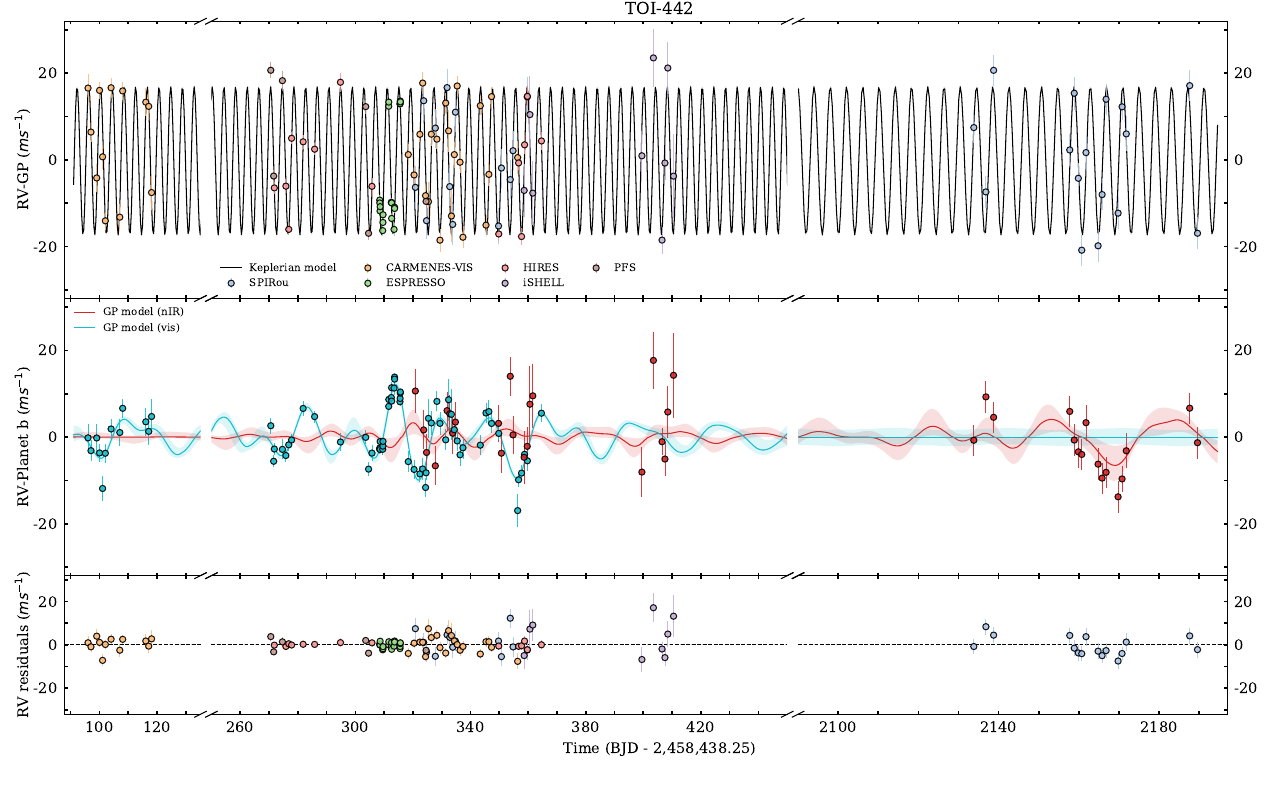}
\caption{Best-fit RV model for TOI-442\,b, the solid black line and its gray overlay correspond to the median and 16th-84th percentile regions of the posterior. Top panel: Keplerian model and drift. Middle panel: dSHO-GP model. Bottom panel: Full model residuals.}
\label{fig:rvfit_442}
\end{figure*}

\begin{figure}
\centering
\includegraphics[width=\hsize]{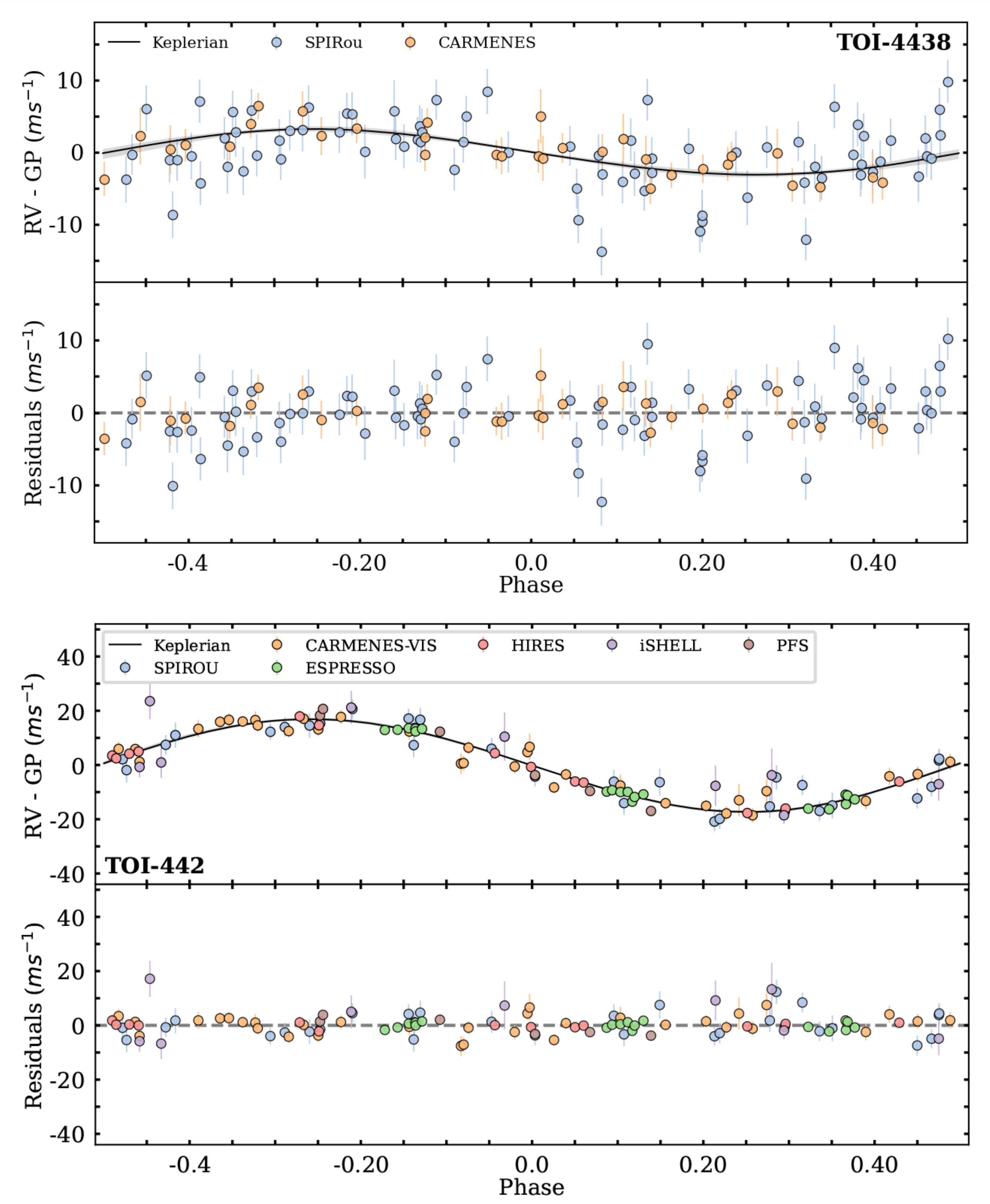}
\caption{Phase folded RVs and residuals. The best-fit model is shown with a solid black line and a gray overlay corresponding to the median and the 16th-84th percentile regions of the posterior. For TOI-4438\,b in the top panel and TOI-442\,b in the bottom panel.}
\label{fig:rvfit_phase}
\end{figure}

\begin{figure}
\centering
\includegraphics[width=\hsize]{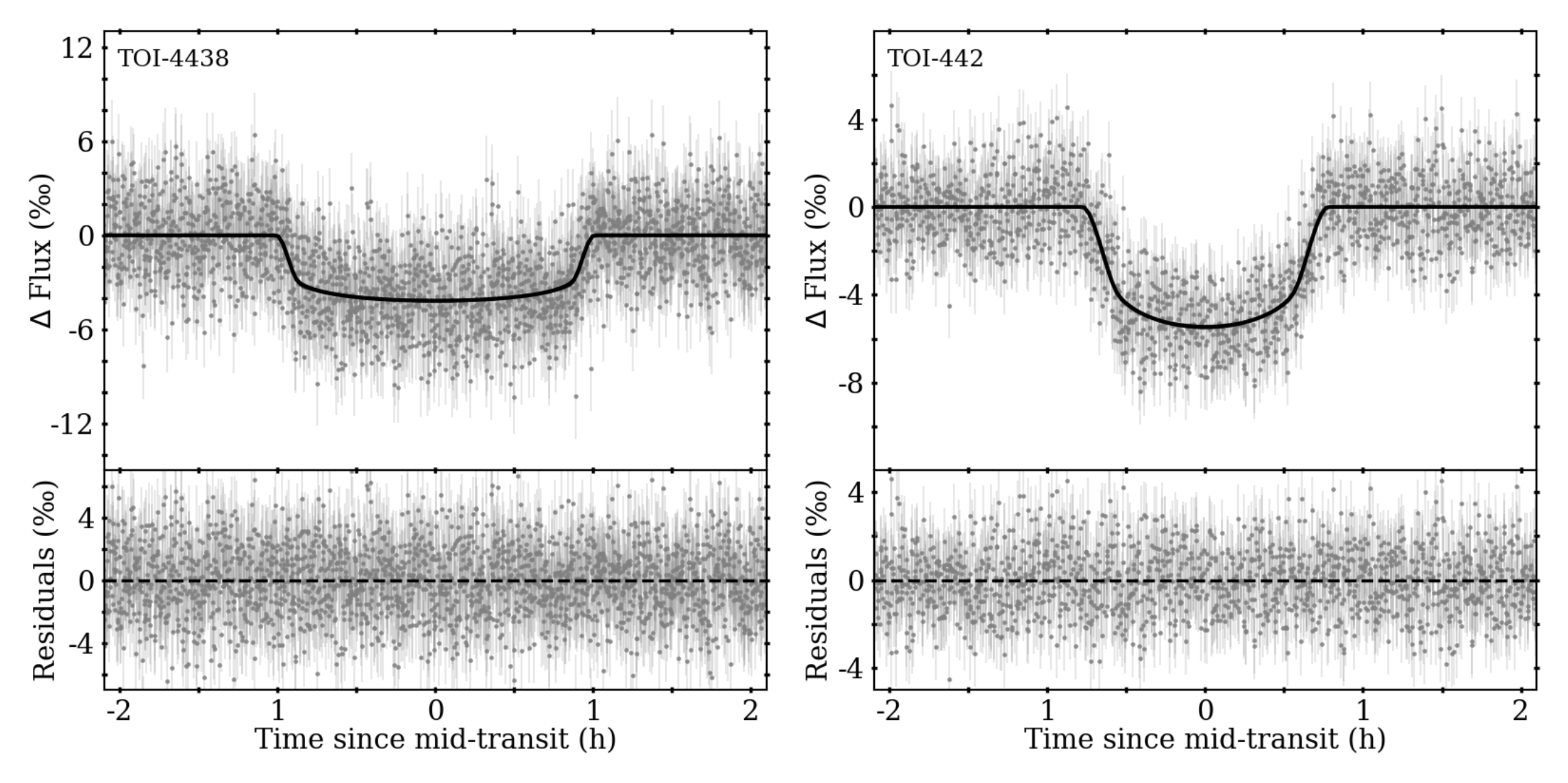}
\caption{Phase folded TESS light curves and residuals. The best-fit model is shown with a solid black line and a gray overlay corresponding to the median and the 16th-84th percentile regions of the posterior. }
\label{fig:lcfit}
\end{figure}

Alternatively, the $\sim$16\,d excess is naturally captured by the dSHO component, and Section~\ref{subsec:periodogram} shows activity power near this period (and near twice this period in both indicators). The prior sensitivity of $P_{\rm c}$, together with the lack of a statistically significant predictive gain of two-Keplerian over GP-only models, argues against the two-planet hypothesis. We therefore cannot rule out an additional planet, but the evidence does not support it at present. We adopt the \textbf{K+D+dSHO} model as the final choice for TOI-442. Figures~\ref{fig:rvfit_442} and \ref{fig:rvfit_phase} show the full and phase-folded RV models, Fig.~\ref{fig:lcfit} shows the transit model, and Table~\ref{tab:results} lists the final system parameters.

\begin{table*}[!t]
\scriptsize
\caption{Priors and results of the joint RV and photometry fit for each system for the chosen final models.}     
\centering     
\begin{tabular}{l c c c c}    
\hline\hline       
 & \multicolumn{2}{c}{TOI-4438\,b} & \multicolumn{2}{c}{TOI-442\,b} \\
\hline      
Parameter & Prior & Posterior & Prior & Posterior \\
\hline
\multicolumn{5}{l}{\textit{Stellar parameters}} \\
    $M_{\star}\,(\msol)$   & B$\mathcal{N}(0.350, 0.018)$ & $0.345^{+0.018}_{-0.018}$ & B$\mathcal{N}(0.542, 0.001)$ & $0.542^{+0.001}_{-0.001}$ \\ 
    $R_{\star}\,(\RS)$  & B$\mathcal{N}(0.355, 0.014)$ & $0.365^{+0.012}_{-0.014}$ &  
    B$\mathcal{N}(0.536, 0.012)$ & $0.536^{+0.001}_{-0.001}$ \\ 
    $q_1$ & $\mathcal{U}(0, 1)$ & $0.23^{+0.25}_{-0.13}$ & 
    $\mathcal{U}(0, 1)$ & $0.51^{+0.28}_{-0.19}$ \\
    $q_2$ & $\mathcal{U}(0, 1)$ & $0.33^{+0.34}_{-0.22}$ & 
    $\mathcal{U}(0, 1)$ & $0.33^{+0.36}_{-0.22}$ \\
\hline
\multicolumn{5}{l}{\textit{Planetary parameters}} \\
    $P$ (days) & $\log \mathcal{N}(2.00772, 8\times10^{-5})$ & $7.446295^{+0.000006}_{-0.000006}$ 
    & $\log \mathcal{N}(1.39923, 8\times10^{-5})$ & $4.052032^{+0.000003}_{-0.000003}$ \\ 
    $T_0$ (TBJD-ref time) & $\log \mathcal{N}(1.7485,0.1)$ & $5.7541^{+0.0006}_{-0.0006}$ & 
    $\log \mathcal{N}(-2.05928,0.1)$ & $0.1316^{+0.0004}_{-0.0004}$ \\ 
    $b$ & $\mathcal{U}(0,1)$ & $0.16^{+0.16}_{-0.11}$ & 
    $\mathcal{U}(0,1)$ & $0.72^{+0.02}_{-0.02}$ \\ 
    $K\,(\ms)$  & $\log \mathcal{N}(1.1726, 0.1)$ & $2.75^{+0.24}_{-0.23}$ & 
    $\log \mathcal{N}(2.9277, 0.1)$ & $17.13^{+0.46}_{-0.44}$ \\ 
    $e$ &  $\mathcal{U}(0,0.5)$ & $0.07^{+0.07}_{-0.05}$ & 
    $\mathcal{U}(0,0.5)$ & $0.022^{+0.022}_{-0.015}$ \\ 
    $\omega$ (rad) & $\mathcal{U}(-\pi,\pi)$ & $-1.5^{+1.5}_{-1.1}$ 
    & $\mathcal{U}(-\pi,\pi)$ & $-2.0^{+3.8}_{-0.5}$ \\ 
    $R_{\rm p}/R_{\star}$  & $\log \mathcal{N}(-2.77650, 0.18317)$ & $0.060^{+0.001}_{-0.001}$ & $\log \mathcal{N}(-2.61723, 0.11620)$ & $0.073^{+0.002}_{-0.001}$ \\ 
    $M_{\rm p}\,(\ME)$  & (derived) & $4.11^{+0.40}_{-0.38}$ 
    & (derived) & $28.38^{+0.77}_{-0.73}$ \\ 
    $R_{\rm p}\,(\RE)$  & (derived) & $2.40^{+0.09}_{-0.10}$  
    & (derived) & $4.25^{+0.10}_{-0.08}$ \\ 
    $a$\,(AU)  & (derived) & $0.0523^{+0.0009}_{-0.0009}$ 
    & (derived) & $0.04056^{+0.00003}_{-0.00003}$  \\ 
    $\rho\,(\gcc)$  & (derived) & $1.65^{+0.26}_{-0.23}$ 
    & (derived) & $2.04^{+0.13}_{-0.14}$ \\ 
    $T_{\text{eq}}~$(K)  & (derived) & $446^{+8}_{-9}$  
    & (derived) & $722.8^{+0.7}_{-0.7}$  \\ 
    $i~$(deg)  & (derived) & $89.7^{+0.2}_{-0.3}$  
    & (derived) & $87.50^{+0.06}_{-0.05}$  \\ 
    $T_{\text{dur}}~$(h)  & (derived) & $1.92^{+0.07}_{-0.10}$ 
    & (derived) & $1.52^{+0.04}_{-0.04}$  \\ 
\hline
\multicolumn{5}{l}{\textit{GP parameters}} \\
    $P_{\rm rot}$  & $\mathcal{N}(77.60, 1.25)$ & $77.7^{+1.3}_{-1.3}$  
    & $\mathcal{N}(28.6, 1.25)$ & $28.8^{+1.5}_{-0.6}$  \\   
    $\sigma_{GP\text{ (nIR)}}$  & H$\mathcal{N}(0,7.5)$ & $4.4^{+1.3}_{-0.9}$ 
    & H$\mathcal{N}(0,13.2)$ & $5^{+3}_{-3}$  \\ 
    $\sigma_{GP\text{ (opt)}}$  & H$\mathcal{N}(0,3.5)$ & $1.2^{+1.2}_{-0.8}$ 
    & H$\mathcal{N}(0,14.9)$ & $6.7^{+2.3}_{-1.4}$  \\
    $\tau_\text{ (nIR)}$  & $\mathcal{U}(10, 1000)$ & $21^{+30}_{-9}$ &  &  \\ 
    $\tau_\text{ (opt)}$  & $\mathcal{U}(10, 1000)$ & $434^{+372}_{-303}$ &  &  \\ 
    $\log {Q_0}_\text{ (nIR)}$  &  &  
    & $\mathcal{U}(1, 8)$ & $2.1^{+2.7}_{-0.8}$ \\ 
    $\log {Q_0}_\text{ (opt)}$  &  & 
    & $\mathcal{U}(1, 8)$ & $1.7^{+0.7}_{-0.5}$ \\ 
    $\log dQ_\text{ (nIR)}$  &  &   
    & $\mathcal{U}(-3, 10)$ & $1.8^{+5.3}_{-3.5}$  \\ 
    $\log dQ_\text{ (opt)}$  &  &   
    & $\mathcal{U}(-3, 10)$ & $6.6^{+2.3}_{-3.5}$  \\ 
    $f_\text{ (nIR)}$  &  &   
    & $\mathcal{U}(0, 1)$ & $0.6^{+0.3}_{-0.3}$ \\ 
    $f_\text{ (opt)}$  &  &   
    & $\mathcal{U}(0, 1)$ & $0.7^{+0.2}_{-0.3}$ \\ 
\hline
\multicolumn{5}{l}{\textit{Instrumental parameters and trend}} \\
    TESS$_{\rm jitter}$ (ppt) & $\log\mathcal{N}(0.91, 1)$ & $0.12^{+0.04}_{-0.05}$  
    & $\log\mathcal{N}(0.61, 1)$ & $0.12^{+0.04}_{-0.05}$\\
    TESS$_{\rm offset}$ (ppt) & $\mathcal{N}(0, 1)$ & $-0.01^{+0.02}_{-0.02}$ 
    & $\mathcal{N}(0, 1)$ & $0.001^{+0.017}_{-0.017}$ \\
    SPIRou$_{\rm jitter}\,(\ms)$  & $\log\mathcal{N}(1.103, 1)$ & $2.8^{+0.5}_{-0.6}$ 
    & $\log\mathcal{N}(1.397, 1)$ & $3.2^{+1.5}_{-2.7}$ \\
    SPIRou$_{\rm offset}\,(\ms)$  & $\mathcal{N}(0, 3)$ & $1.6^{+1.6}_{-1.6}$  
    & $\mathcal{N}(0, 5)$ & $-1.2^{+1.7}_{-1.6}$ \\
    CARMENES-VIS$_{\rm jitter}\,(\ms)$  & $\log\mathcal{N}(0.835, 1)$ & $1.0^{+0.5}_{-0.7}$ 
    & $\log\mathcal{N}(0.947, 1)$ & $2.3^{+0.8}_{-1.1}$  \\
    CARMENES-VIS$_{\rm offset}\,(\ms)$  & $\mathcal{N}(0, 3)$ & $0.17^{+1.4}_{-1.4}$ 
    & $\mathcal{N}(0, 5)$ & $-2.1^{+1.0}_{-1.0}$  \\
    ESPRESSO$_{\rm jitter}\,(\ms)$  & & 
    & $\log\mathcal{N}(-0.051, 1)$ & $1.0^{+0.3}_{-0.4}$ \\
    ESPRESSO$_{\rm offset}\,(\ms)$  & &  
    & $\mathcal{N}(0, 5)$ & $12.3^{+1.7}_{-1.7}$ \\
    HIRES$_{\rm jitter}\,(\ms)$  & & 
    & $\log\mathcal{N}(0.506, 1)$ & $0.7^{+0.4}_{-0.5}$ \\
    HIRES$_{\rm offset}\,(\ms)$  & & 
    & $\mathcal{N}(0, 5)$ & $2.3^{+0.9}_{-0.9}$ \\
    iSHELL$_{\rm jitter}\,(\ms)$  & & 
    & $\log\mathcal{N}(1.813, 1)$ & $5.5^{+2.8}_{-3.8}$ \\
    iSHELL$_{\rm offset}\,(\ms)$  & & 
    & $\mathcal{N}(0, 5)$ & $4.6^{+2.8}_{-2.9}$ \\
    PFS$_{\rm jitter}\,(\ms)$  & & 
    & $\log\mathcal{N}(0.479, 1)$ & $2.7^{+1.2}_{-1.4}$ \\
    PFS$_{\rm offset}\,(\ms)$  & & 
    & $\mathcal{N}(0, 5)$ & $-6.9^{+1.8}_{-1.7}$ \\
    RV trend C$_{0}$ & $\mathcal{N}(0, 1)$ & $0.2^{+0.9}_{-0.9}$  & & \\
    RV trend C$_{1}$ & $\mathcal{N}(0, 0.1)$ & $-0.011^{+0.004}_{-0.004}$ & &  \\
    RV trend C$_{2}$ & $\mathcal{N}(0, 0.01)$ & $-0.00004^{+0.00001}_{-0.00001}$ & &  \\
\hline
\label{tab:results}
\end{tabular}
\tablefoot{$\mathcal{N}(\mu, \sigma)$ stands for normal distribution, in the case of stellar mass and radius B$\mathcal{N}$ means bounded normal and both distributions are bounded between 0 and 1. $\mathcal{U}(a,b)$, $\log \mathcal{N}(\mu,\sigma)$ to a log-normal distribution and H$\mathcal{N}(\mu, \sigma)$ is the half-normal distribution. Posterior values are medians with 16th--84th percentile uncertainties. The transit duration $T_{\text{dur}}~$ is calculated with the expression given by \cite{winn2014transitsoccultations} for the total duration (from the first contact to the fourth). The reference time for $T_0$ is $2390.65659$\,TBJD for TOI-4438\,b and $1438.25327$\,TBJD for TOI-442\,b.}
\end{table*}

\section{Discussion}
\label{sec:discussion}
In our analysis of the TOI-4438 system, we find that the mini-Neptune TOI-4438\,b has a mass of $4.11^{+0.40}_{-0.38}\,\ME$, a radius of $2.40^{+0.09}_{-0.10}\,\RE$ and an eccentricity of $0.07^{+0.07}_{-0.05}$. Our results improve the relative uncertainties in mass and radius from 20.3\% to 9.7\% and 5.1\% to 4.0\% respectively, and imply a bulk density of $1.65^{+0.26}_{-0.23}\,\gcc$. The values are consistent with those found by \citet{Goffo2024} within 1$\sigma$. 

\begin{figure}[!t]
\centering
\includegraphics[width=\hsize]{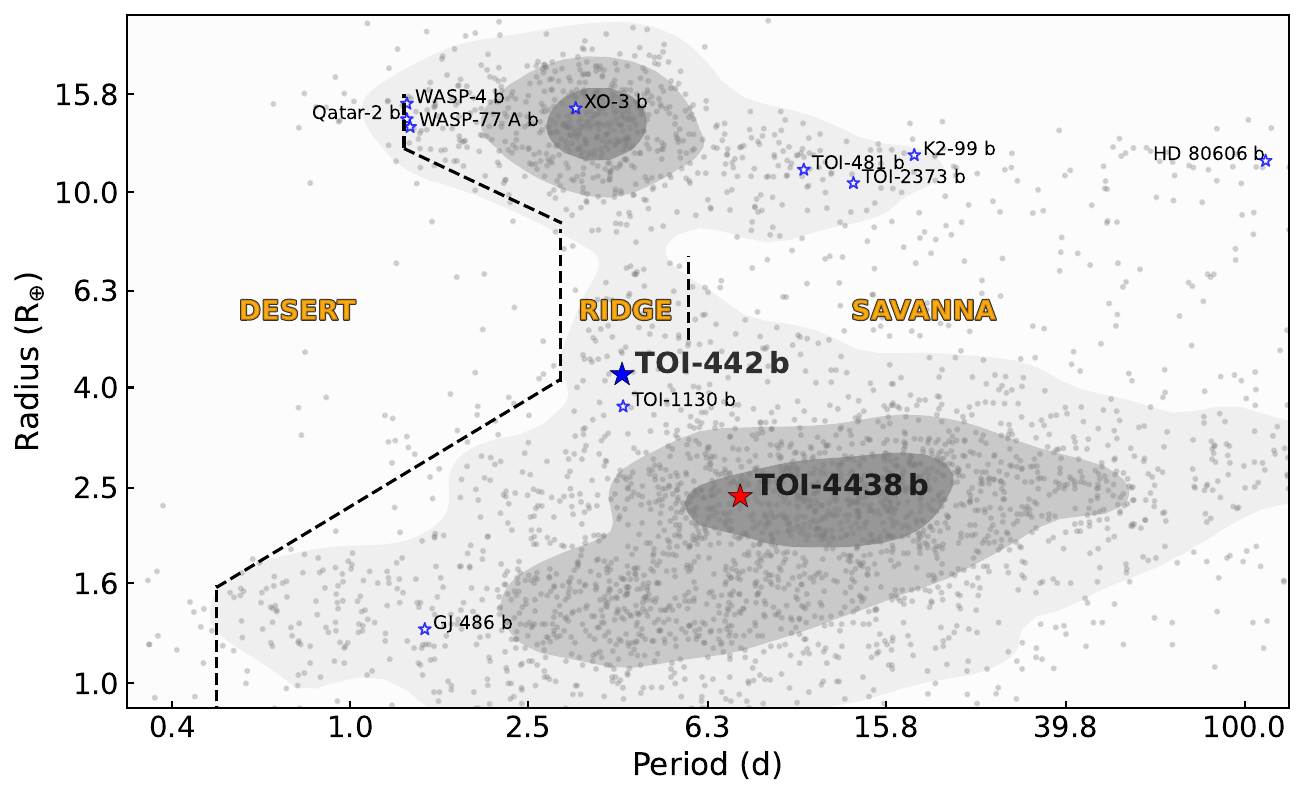}
\caption{Radius vs orbital period of known exoplanets obtained from the NASA Exoplanet Archive (planets with radius precision greater than 20\%). Dashed black lines mark the estimated boundaries for the hot-Neptune "desert", "ridge" and "savanna" \citep{Castro-Gonzalez2024}. The position of TOI-4438\,b and TOI-442\,b is marked with red and blue markers respectively. Planets with mass and radius measured with equal or more precision than TOI-442\,b are marked with outlined blue markers and names.}
\label{fig:desert}
\end{figure}

For TOI-442, we derive for the hot Neptune TOI-442\,b a mass of 28.38$^{+0.77}_{-0.73}\,\ME$ and a radius of 4.25$^{+0.10}_{-0.08}\,\RE$, which are compatible within 1.4$\sigma$ from the values found by \citet{Dreizler2020}, but reduce the relative uncertainties from 4.8\% to 2.7\% and from 6.3\% to 2.3\%, respectively. This results in a bulk density of $2.04^{+0.13}_{-0.14}\,\gcc$. We also find the planet to be in a nearly-circular orbit, with $e=0.02^{+0.02}_{-0.02}$. The remarkable precision on the bulk parameters of TOI-442\,b places it among the $\sim$10 planets with the most precisely measured masses and radii known to date\footnote{Source: \url{https://exoplanetarchive.ipac.caltech.edu/}}, and the most precise from the Neptunian $(4\,\RE\,<\,R_{\rm p}\,<\,10\,\RE)$ planets (see Fig. \ref{fig:desert}).

Notably, the chromatic behavior appears to differ between the two systems, although the significance of the differences is modest.  For TOI-4438, the nIR GP amplitude ($4.4^{+1.3}_{-0.9}\,\ms$) appears larger than the optical ($1.2^{+1.2}_{-0.8}\,\ms$), though the two are statistically  consistent within $\sim2\sigma$. While a stronger nIR activity signal may seem counterintuitive from a spot temperature-contrast perspective (spot contrast decreases toward longer wavelengths; \citealt{Reiners2010}) it is a predicted consequence of Zeeman broadening: when magnetic activity is significant, the Zeeman-induced RV signal grows with wavelength and can overcome the spot-contrast contribution \citep{Reiners2013, Hebrard2014}. This is consistent with the polarimetric analysis of TOI-4438 in Section~\ref{subsec:stellaractivity}, which reveals the presence of magnetic active regions whose signatures vary in time, indicative of a complex or moderately strong magnetic field configuration. In such a regime, Zeeman broadening may dominate the nIR RV variability. We also note that residual telluric contamination in SPIRou spectra is a known contributor to inflated nIR RV scatter \citep[e.g.,][]{Ould-Elhkim2026}, and may further contribute to the elevated nIR amplitude. For TOI-442, the optical ($6.7^{+2.3}_{-1.4}\,\ms$) and nIR ($5\pm3\,\ms$) amplitudes are consistent within their uncertainties, in line with theoretical expectations for an early M dwarf \citep{Reiners2010} and with previous observational results for stars of similar spectral type \citep{Cortes-Zuleta2023}. The strongly and consistently detected Zeeman signatures in TOI-442  (Section~\ref{subsec:stellaractivity}), are consistent with a scenario in which both spot-contrast and Zeeman broadening contribute to the RV variability, which could explain the comparable amplitudes observed  across optical and nIR wavelengths.

We find that including a quadratic drift does not substantially alter the inferred planetary parameters in either model. Nevertheless, for TOI-4438 in all cases the LOO-CV metric shows a marginal improvement within the associated uncertainties when the drift term is included. The final model chosen includes a drift, and both the quadratic and linear coefficients have a 95\% HDI that excludes zero, indicating statistically significant curvature in the RV time series. The inferred amplitudes of this curvature are $12\pm3\,\ms$ over the observational baseline for TOI-4438. We found no statistical evidence for long-term drift in the case of TOI-442.

We refine the Transmission and Emission Spectroscopy Metrics \citep[TSM,ESM;][]{Kempton2018} for both planets. For TOI-4438\,b we obtain $\mathrm{TSM}=163^{+27}_{-28}$ and $\mathrm{ESM}=3.9\pm0.5$, while for TOI-442\,b we find $\mathrm{TSM}=98^{+8}_{-6}$ and $\mathrm{ESM}=24\pm1$. Both planets exceed the $\mathrm{TSM} \gtrsim 90$ threshold for high-quality transmission spectroscopy targets with \textit{JWST}, with TOI-4438\,b ranking particularly high. In contrast, only TOI-442\,b exceeds the suggested $\mathrm{ESM}$ threshold of 7.5 \citep{Kempton2018}, making it a favorable target for emission spectroscopy. Both planets are also included in the ARIEL Mission Candidate Sample (MCS\footnote{Source: \url{https://github.com/arielmission-space/Mission_Candidate_Sample}}).

\subsection{Location of TOI-442\,b in the Neptune desert}
\label{subsec:recharacterization_toi442}
TOI-442\,b occupies a region of low occurrence in both the period–mass and period–radius distributions, corresponding to Neptune-size planets with orbital periods shorter than 30 days. Its initial characterization identified TOI-442\,b as a warm Neptune-like planet located near the boundary of the hot Neptune desert as defined by \cite{Mazeh2016}. Planets located within or near the boundaries of the hot Neptune desert are key targets for constraining the physical processes that shape this region of parameter space. Several mechanisms have been proposed to drive planetary evolution in this regime. Atmospheric mass loss from hydrodynamical expansion and escape, and photoevaporation in particular, can erode the gaseous envelopes of Neptune-size planets, transforming them into sub-Neptunes or exposing their rocky cores \citep[e.g.,][]{Lecavelier2007, Murray-Clay2009, Lopez2013, Owen2017}. In addition, dynamical processes such as high eccentricity migration (HEM) and disk-driven migration may deliver giant planets to close-in orbits, where subsequent atmospheric loss further alters their properties. The relative importance of these mechanisms likely depends on the evolutionary stage of the planetary system. 

At longer periods, the desert transitions into the more populated Neptunian savanna \citep{Bourrier2023}. \citet{Castro-Gonzalez2024} place the upper period limit of the desert at $\approx$3.2~days, with an overdensity of planets between $3.2 \lesssim P \lesssim 5.7$~days and radii $5.5$–$8.5\,\RE$, referred to as the Neptunian ridge. With $R_{\rm p}=4.25^{+0.10}_{-0.08}\,\RE$ and $P=4.05$~days, TOI-442~b lies near the desert–savanna transition but outside the ridge due to its smaller radius. Most warm Neptunes $(P\lessapprox5\,\rm{d})$ are found to have non-zero eccentricities \citep{Correia2020}, supporting the HEM process for populating the desert. The orbit of TOI-442~b appears to be almost circular, and whether or not it underwent atmospheric loss remains an open question for future studies.

\subsection{Interior modeling}
\label{sec:interior}

We analyzed the potential composition of TOI-4438\,b and TOI-442\,b using the inference model of \citet{DeWringer2026} with interior models based on \cite{Dorn2015, Dorn2017, luo_majority_2024}. The underlying forward model consists of three layers: an iron core, a silicate mantle, and a H$_2$-He-H$_2$O atmosphere. A detailed description of the forward model and inference framework is provided in Appendix~\ref{app:interior}.

The interior model is optimized for super-Earths and sub-Neptunes with atmospheric mass fractions $\lesssim15\%$. TOI-442\,b lies at the upper end of this range; the model grid covers its parameters, so the inference remains valid, but we account for the larger atmospheric contribution by imposing a prior on $M_{\rm core+mantle}$ at 90\% of the total mass with an inflated uncertainty.

The planetary mass, radius, and equilibrium temperature are taken from the joint transit and radial-velocity analysis (Table~\ref{tab:results}), and the stellar age from Table~\ref{tab:stellar_params}. These are treated as observational constraints. The prior distributions and inference results are summarized in Table~\ref{tab:interior_combined} and Figures~\ref{fig:surrogate_posteriors_4438} and~\ref{fig:surrogate_posteriors_442}. Consistent with their low bulk densities, we infer volatile-rich compositions with substantial atmospheres. For TOI-4438\,b, we find an atmospheric mass fraction of $7_{-2}^{+3}$\%, while TOI-442\,b has a larger value of $18_{-3}^{+5}$\%. The inferred atmospheric metallicities are $Z_{\rm env} = 0.30_{-0.17}^{+0.14}$ and $Z_{\rm env} = 0.22^{+0.15}_{-0.14}$, respectively, both consistent with super-solar values but weakly constrained.

Our results can be compared with previous interior analyses of these systems. For TOI-442\,b, no quantitative interior inference exists in the literature: \citet{Dreizler2020} placed the planet on a mass--radius diagram against the composition models of \citet{Zeng2016}, arguing qualitatively from its low bulk density that it retains a significantly thick gaseous envelope. Our analysis therefore provides the first characterization of its interior. For TOI-4438\,b, \citet{Goffo2024} modeled the planet with a three-layer structure \citep[Fe core, silicate mantle, and a pure-water envelope;][]{Acuna2021,Aguichine2021}, setting the iron fraction to zero and finding a water mass fraction of $0.62^{+0.34}_{-0.16}$. A direct quantitative comparison with our value is not meaningful, since the two analyses adopt different envelope compositions: \citet{Goffo2024} explore a water-dominated end-member, whereas our model assumes an H$_2$-He-H$_2$O envelope on an Earth-like rocky interior. Both analyses nonetheless agree that the planet is volatile-rich and of low density, and both are consistent with the mass--composition degeneracy noted by \citet{Goffo2024}, who could not distinguish a massive, high-metallicity (water-rich) envelope from a less massive, low-metallicity (H/He-rich) one without atmospheric data.

These atmospheric mass fractions are lower bounds: our model confines hydrogen to the envelope, but hydrogen and silicate melt are miscible at sub-Neptune interior conditions \citep{Young2024, Rogers2025}, so part of the hydrogen can dissolve into the deep interior, an effect that grows over the planet's history as interior hydrogen later exsolves \citep{Rogers2025}. 

TOI-442\,b's position at the bottom of the Neptunian ridge (R$_p$ = 4.25\,R$_\oplus$, P = 4.05\,d) may reflect sculpting by atmospheric escape (Section~\ref{subsec:recharacterization_toi442}). Its current radius lies far below the core-powered mass-loss---photoevaporation transition radius predicted by \citet{misener_characterizing_2026} (using their Eq. 11, R$_\mathrm{trans}\sim$ 24 R$_\oplus$ for this planet's mass and T$_{\rm eq}$) indicating it has long since transitioned into the photoevaporative regime, where UV-driven escape has been operating potentially since early in the system's history. Given the host star's M-dwarf spectral type, XUV luminosity is expected to be moderate compared to solar-type stars, which may have slowed photoevaporative stripping relative to a solar analog. The nearly circular orbit (e $\sim$ 0.02) is also consistent with a planet that did not undergo significant high-eccentricity migration, suggesting in-situ or disk-driven migration followed by gradual atmospheric erosion rather than dynamical delivery into the desert. Whether the current $18_{-3}^{+5}$\% atmospheric mass fraction represents a remnant envelope that survived photoevaporation, or whether the planet is still losing atmosphere today, remains an open question; transmission spectroscopy could simultaneously help address this and break the envelope composition degeneracy by directly constraining the metallicity.

\section{Conclusions}
\label{sec:conclusion}
We provided a comprehensive re-analysis of the TOI-4438 and TOI-442 systems by integrating new TESS photometry and high-resolution spectropolarimetry from SPIRou with archival data. The analysis allowed for a significant refinement of bulk planetary properties, reducing the mass uncertainties by 53\% and 46\% respectively for TOI-4438\,b and TOI-442\,b and the radii by 22\% and 67\%. The tighter constraints result from the enlarged photometric and RV data sets for both systems, with additional gains for TOI-442 from explicitly modeling stellar-activity variability in the RVs. 

A search for additional planets did not reveal significant signals in either system. In TOI-4438, a single transit-like feature is present in the TESS light curve but cannot be confirmed as planetary with current data; we constrain its period to $74^{+152}_{-34}$\,d and radius to $1.8 \pm 0.1\,R_{\oplus}$. Additional TESS observations, long-baseline photometry, or higher-precision RVs are needed to test its planetary nature. For TOI-442, the RV variability near 16\,d is more naturally explained by stellar activity and is well captured by a quasi-periodic Gaussian process, which improves the fit and tightens the constraints on TOI-442\,b. We find no compelling evidence for an additional planet in the current data.

We modeled the internal structure of both planets, constraining the presence of gaseous envelopes. TOI-4438\,b is compatible with a modest atmospheric mass fraction, while TOI-442\,b requires a substantially more massive atmosphere. Although the envelope metallicities are weakly constrained, the inferred interior structures remain consistent with super-solar values. The improved precision on the planetary masses and radii obtained in this work provides a favorable basis for future atmospheric characterization, and both planets are suitable targets for follow-up atmospheric studies.

\begin{acknowledgements}
E.M. acknowledges funding from FAPEMIG under project number APQ-02493-22 and a research productivity grant number 309829/2022-4 awarded by the CNPq.

This research has made use of the NASA Exoplanet Archive, which is operated by the California Institute of Technology, under contract with the National Aeronautics and Space Administration under the Exoplanet Exploration Program. 

C.D. acknowledges support from the Swiss National Science Foundation under grant TMSGI2\_211313 and the COPL grant \textit{Evolution and Diversity of Super-Earth Atmospheres}. This work has been carried out within the framework of the NCCR PlanetS supported by the Swiss National Science Foundation under grant 51NF40\_205606.

This paper made use of data collected by the TESS mission and are publicly available from the Mikulski Archive for Space Telescopes (MAST) operated by the Space Telescope Science Institute (STScI). Funding for the TESS mission is provided by NASA’s Science Mission Directorate. We acknowledge the use of public TESS data from pipelines at the TESS Science Office and at the TESS Science Processing Operations Center. Resources supporting this work were provided by the NASA High-End Computing (HEC) Program through the NASA Advanced Supercomputing (NAS) Division at Ames Research Center for the production of the SPOC data products.

A.L'H. acknowledges financial support from the Fonds de recherche du Québec - Secteur Nature et technologies (FRQNT) under file \#349961.
\end{acknowledgements}

\bibliographystyle{aa} 
\bibliography{aa} 

\appendix
\section{Additional figures and tables}
\label{apx:figs}

\begin{figure}
\centering
   \includegraphics[width=0.9\hsize]{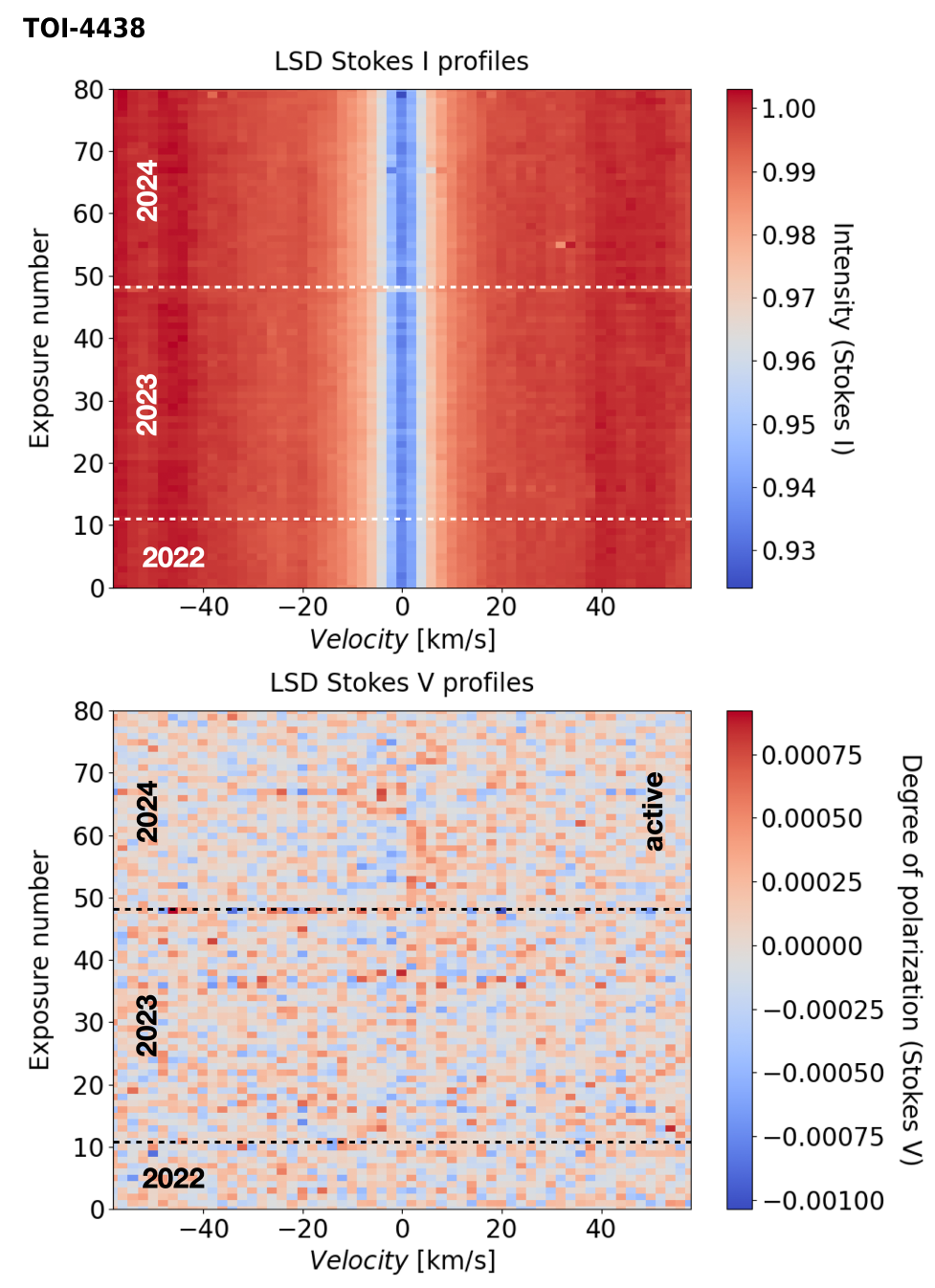}
    \caption{Stokes~I (top) and Stokes~V (bottom) profiles from the LSD analysis of the SPIRou spectropolarimetric observations of TOI-4438.}
    \label{fig:lsd_img2_4438}
\end{figure}

\begin{figure}
\centering
   \includegraphics[width=0.9\hsize]{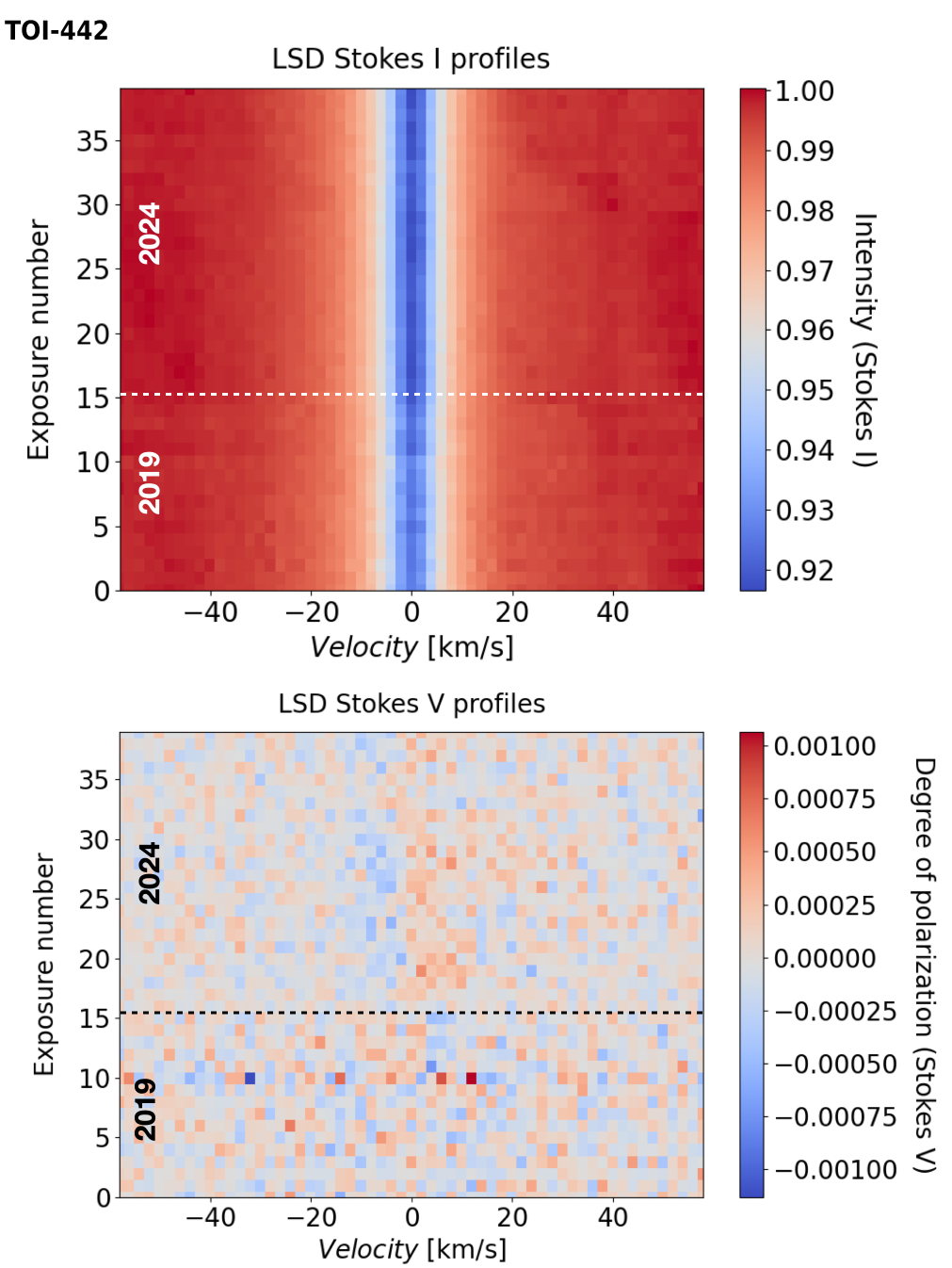}
    \caption{Stokes~I (top) and Stokes~V (bottom) profiles from the LSD analysis of the SPIRou spectropolarimetric observations of TOI-442.}
    \label{fig:lsd_img2_442}
\end{figure}

\begin{table}
\caption{Priors and best-fit posterior values of the GP model hyperparameters obtained from our analysis of the stellar activity in the SPIRou $B_\ell$ data.}
\footnotesize
\label{tab:activitygpfitparams}
\begin{center}
\begin{tabular}{lccc}
    \hline
    \hline
    \noalign{\smallskip}
Quantity & Priors & \multicolumn{2}{c}{ Fit values }\\
&  & TOI-4438 & TOI-442 \\
    \noalign{\smallskip}
    \hline
$\mu$ (G) & $\mathcal{U}(-\infty,+\infty)$ & $-3\pm8$ & $-5\pm5$ \\
$\sigma$ (G) & $\mathcal{U}(0,+\infty)$ & $3_{-2}^{+3}$   & $3_{-2}^{+3}$ \\
$\alpha$ (G) & $\mathcal{U}(0,+\infty)$ &  $16_{-4}^{+7}$ & $10_{-4}^{+5}$ \\
$l$ (d) & $\mathcal{U}(10,1000)$ & $1216_{-880}^{+1150}$ & $484_{-310}^{+344}$ \\
$\beta$ & Fixed  & $0.5$  & $0.5$ \\
$P_{\rm rot}$ (d) & $\mathcal{U}(2,300)$ &  $77.6^{+0.9}_{-1.6}$ & $28.6^{+1.6}_{-0.9}$ \\
Residuals RMS [G] & & 14.7 & 8.8 \\
$\chi^2$ & & 0.93 & 0.84 \\
        \hline
    \end{tabular}
    \end{center}
\tablefoot{$\mathcal{U}(a,b)$ refers to a uniform distribution. The parameter $\mu$ represents the mean, $\sigma$ the white noise, $\alpha$ the amplitude, $l$ the decay time, $\beta$ is the smoothing factor and $P_{\rm rot}$ the rotation period. The reported fit values correspond to the median and the 34th and 86th percentiles.}
\end{table}

\begin{figure}
\centering
\includegraphics[width=0.9\hsize]{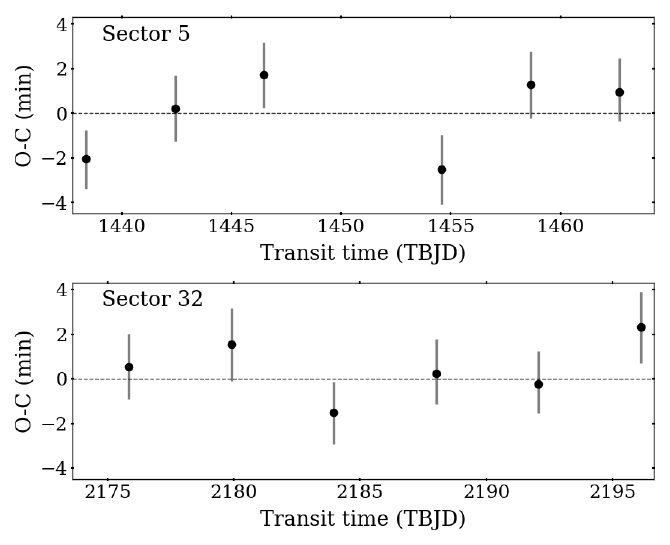}
\caption{Results of the transit timing variations model for the TESS light curve of TOI-442\,b. The O-C values represent the difference between the fitted value for each mid-transit time and the calculated value considering a fixed period and reference mid-transit.}
\label{fig:ttvs}
\end{figure}

\begin{table}
\caption{Priors and posterior parameters for the single–transit candidate in the TOI-4438 system.}
\label{tab:single_transit}
\centering
\footnotesize
\begin{tabular}{l c c}
\hline\hline
Parameter & Prior & Posterior \\
\hline
$M_\star$\,($M_\odot$) & $\mathcal{N}(0.345,\,0.017)$ & $0.345^{+0.017}_{-0.017}$ \\
$R_\star$\,($R_\odot$) & $\mathcal{N}(0.363,\,0.013)$ & $0.363^{+0.014}_{-0.013}$ \\
$t_0$\,(TBJD) & $\mathcal{N}(3471.559,\,0.1)$ & $3471.566^{+0.009}_{-0.006}$ \\
$P$\,(d) & log$\mathcal{U}(32.7, 2000)$ & $74^{+152}_{-34}$ \\
$b$ & $\mathcal{U}(0,\,1)$ & $0.8^{+0.1}_{-0.2}$ \\
$R_{\rm p}/R_\star$ & $\mathcal{N}(0.045,\,0.0032)$ & $0.045^{+0.003}_{-0.003}$ \\
$\rho_\star$\,(g\,cm$^{-3}$) & Derived & $10^{+1}_{-1}$ \\
$a/R_\star$ & Derived & $143^{+157}_{-47}$ \\
$T_{14}$\,(h) & Derived & $2.9^{+0.6}_{-0.3}$ \\
$R_{\rm p}$\,($R_\oplus$) & Derived & $1.8\pm0.1$ \\
$a$\,(AU) & Derived & $0.24^{+0.27}_{-0.08}$ \\
$T_{\rm eq}$\,(K) & Derived & $207^{+46}_{-64}$ \\
\hline
\end{tabular}
\end{table}

\begin{figure}
\centering
\includegraphics[width=0.9\hsize]{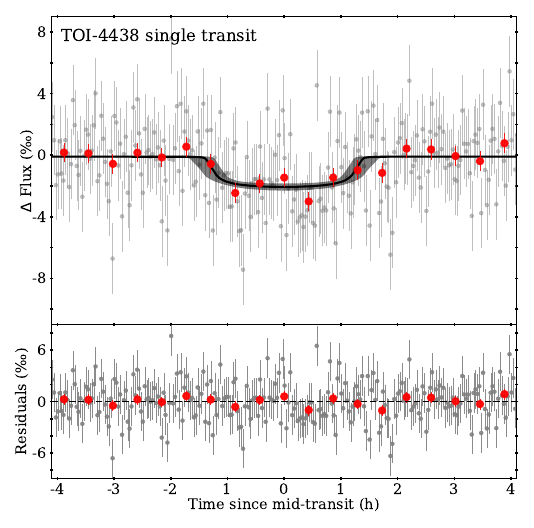}
\caption{Phase folded TESS light curve and residuals for the single-transit event detected for TOI-4438. The unbinned photometric data are shown as gray points, while red points indicate the light curve binned in time for visualization. The best-fit transit model is shown with a solid black line, with the shaded region corresponding to the 16th--84th percentile range of the posterior distribution.}
\label{fig:single_transit}
\end{figure}

\begin{table}[!h]
\caption{Posterior parameters for the candidate second planet in the TOI-442 system, shown for the different models.}         
\label{table:planetc}      
\centering     
\footnotesize                           
\begin{tabular}{l c c c }          
\hline\hline                        
Parameter  & Prior & 2K+D & 2K+D+SHO \\    
\hline               
$K_{\rm c}\,(\ms)$ & $\mathcal{HN}(0, 5)$ & $6.8^{+0.4}_{-0.4}$ & $5.8^{+0.7}_{-0.8}$ \\
$P_{\rm c}$\,(d) & $\mathcal{U}(14, 18)$ & $16.029^{+0.009}_{-0.017}$ & $16.00^{+0.02}_{-0.01}$ \\
$\lambda_{0 \rm c}$ & $\mathcal{U}(0, 2\pi)$ & $3.3^{+2.1}_{-2.3}$ & $2.9^{+2.5}_{-2.4}$ \\ 
$t_{\rm peri\,c}$\,(TBJD) & Derived & $2022.7^{+0.6}_{-0.8}$ & $2022.4^{+0.9}_{-1.0}$ \\
$M_{\rm c}\sin{i}\,(\ME)$ & Derived & $18^{+1}_{-1}$ & $15^{+2}_{-2}$ \\
$e_{\rm c}$ & $\mathcal{U}(0, 0.7)$ & $0.33^{+0.07}_{-0.07}$ & $0.27^{+0.09}_{-0.10}$ \\
$\omega_{\rm c}$ & $\mathcal{U}(0, 2\pi)$ & $5.2^{+0.3}_{-0.3}$ & $5.2^{+0.4}_{-0.4}$ \\
\hline                                             
\end{tabular}
\end{table}

\begin{figure}
\centering
\includegraphics[width=\hsize]{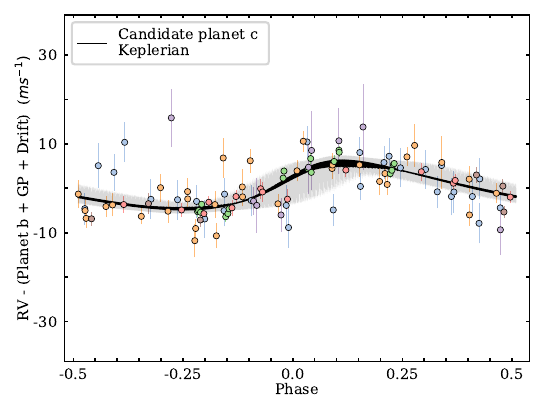}
\caption{Phase folded RVs for TOI-442 after subtracting the signals from planet TOI-442\,b, the SHO GP and the quadratic drift. The best-fit Keplerian model for the candidate planet c is shown in black with the shaded region corresponding to the 16th--84th percentile range of the posterior distribution. The model corresponds to the 2K+SHO+D version.}
\label{fig:planetc_fit}
\end{figure}

\begin{figure}
\centering
\includegraphics[width=0.9\hsize]{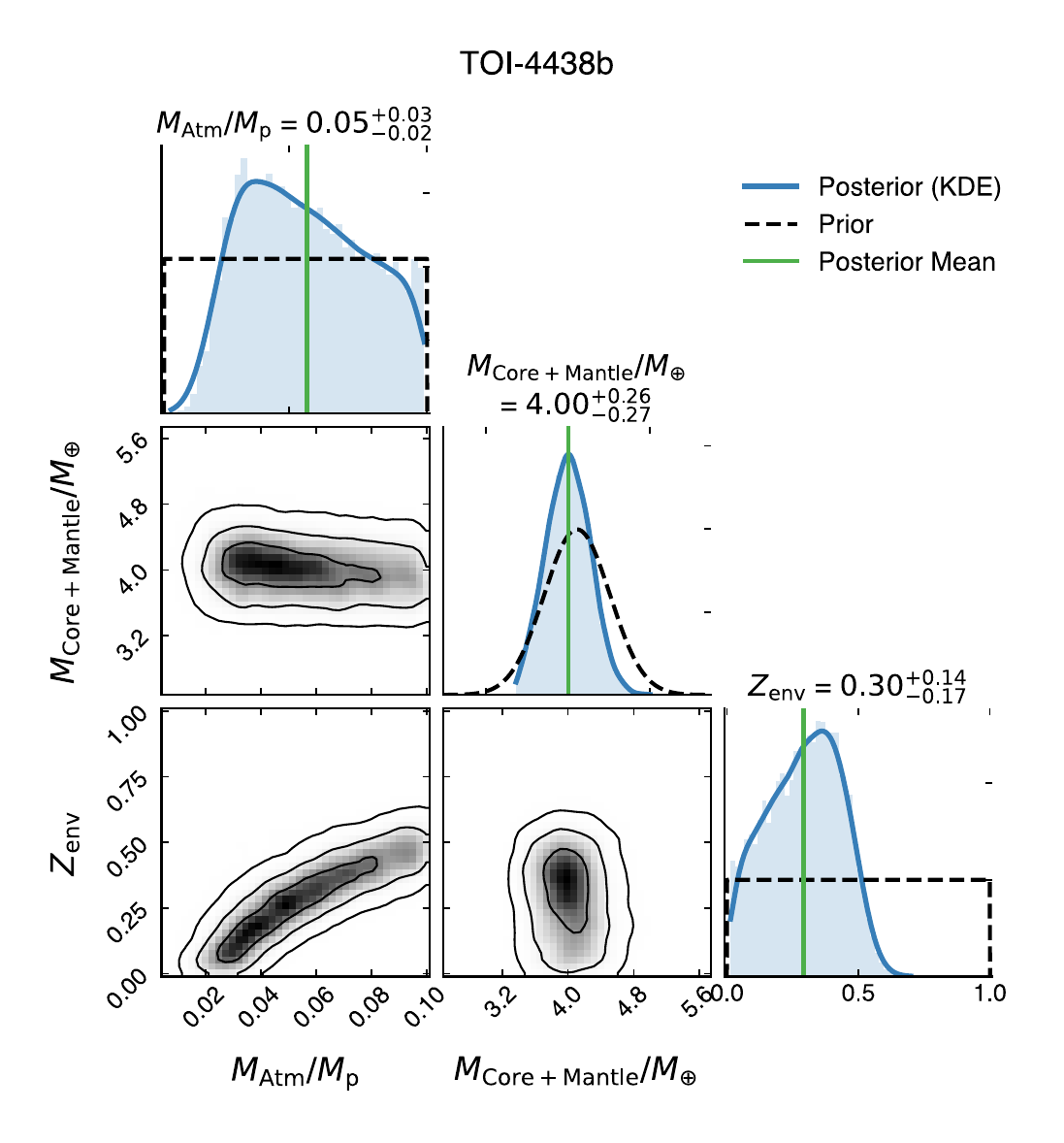}
\caption{
Posterior distributions and parameter correlations for TOI-4438 b from the interior model. Diagonal panels show one-dimensional marginal posterior distributions for the atmospheric mass fraction ($M_{\mathrm{atm}}/M_{\mathrm{tot}}$), core+mantle mass ($M_{\mathrm{core+mantle}}$), and envelope metallicity ($Z_{\mathrm{env}}$). Solid blue curves represent kernel density estimates of the posterior, while dashed black lines indicate the corresponding prior distributions. Vertical green lines mark posterior mean values, with annotated uncertainties indicating the central credible intervals. Off-diagonal panels display two-dimensional posterior density contours, illustrating correlations between inferred parameters.
}
\label{fig:surrogate_posteriors_4438} 
\end{figure}

\begin{figure}
\centering
\includegraphics[width=0.9\hsize]{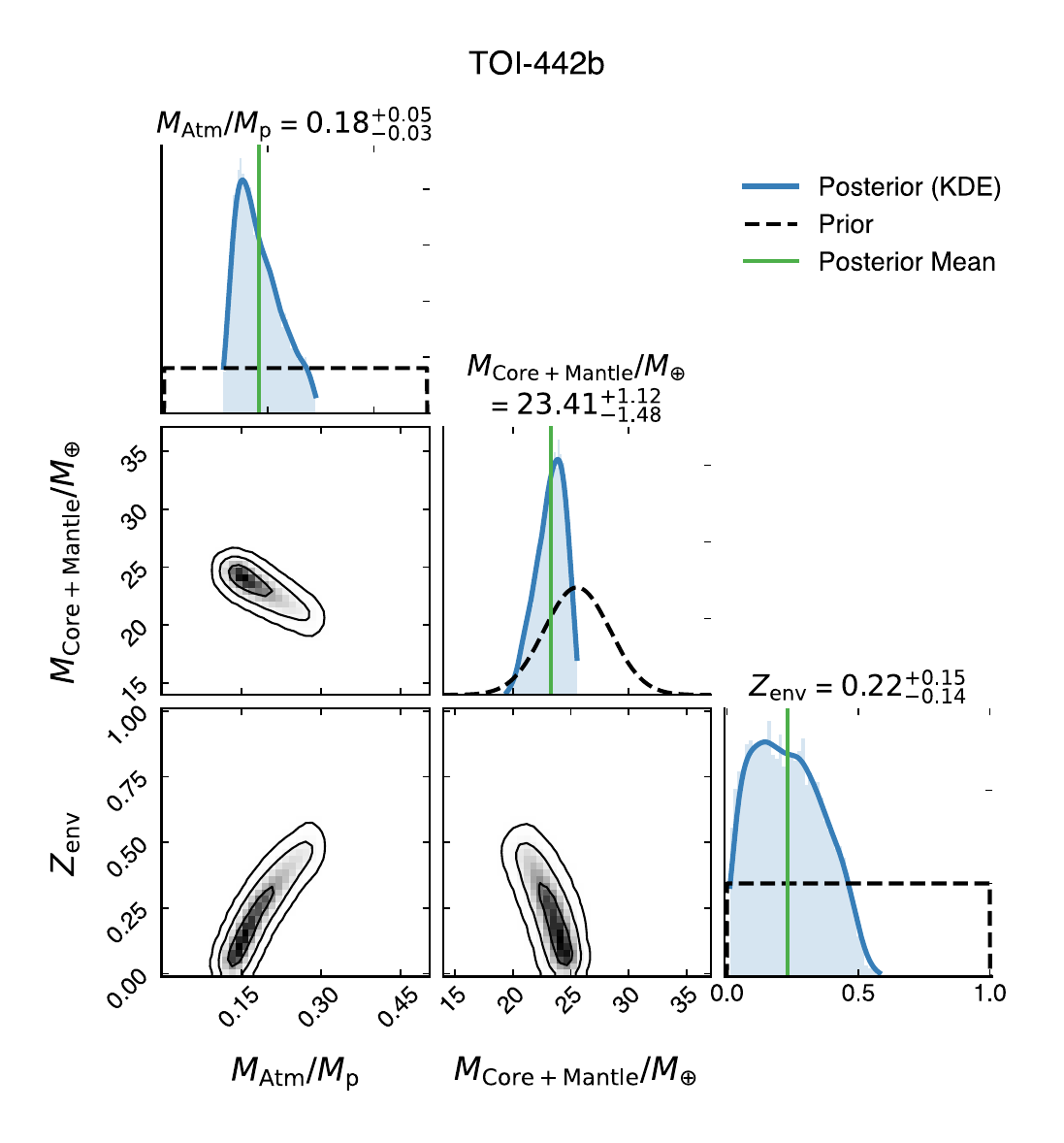}
\caption{
Posterior distributions and parameter correlations for TOI-442 b from the interior model. Diagonal panels show one-dimensional marginal posterior distributions for the atmospheric mass fraction ($M_{\mathrm{atm}}/M_{\mathrm{tot}}$), core+mantle mass ($M_{\mathrm{core+mantle}}$), and envelope metallicity ($Z_{\mathrm{env}}$). Solid blue curves represent kernel density estimates of the posterior, while dashed black lines indicate the corresponding prior distributions. Vertical green lines mark posterior mean values, with annotated uncertainties indicating the central credible intervals. Off-diagonal panels display two-dimensional posterior density contours, illustrating correlations between inferred parameters.
}
\label{fig:surrogate_posteriors_442} 
\end{figure}

\begin{table}[!htbp]
\centering
\footnotesize
\caption{Prior distributions and posterior results for the interior composition of TOI-4438\,b and TOI-442\,b. Reported uncertainties correspond to $1\sigma$ credible intervals. We report the atmospheric mass fraction and the total mass of the rocky interior (core+mantle).}
\label{tab:interior_combined}
\begin{tabular}{lcc}
\hline\hline
Parameter & TOI-4438\,b & TOI-442\,b \\
\hline
$M_{\rm atm}/M_\text{p}$ prior & ${\rm log\,} \mathcal{U}(0.005,0.10)$ & ${\rm log\,} \mathcal{U}(0.05,0.50)$ \\
$M_{\rm atm}/M_\text{p}$ posterior & $0.05^{+0.03}_{-0.02}$ & $0.18^{+0.05}_{-0.03}$ \\
$M_{\rm core+mantle}/M_{\oplus}$ prior & $\mathcal{N}(4.11,0.40)$ & $\mathcal{N}(25.54,2.84)$ \\
$M_{\rm core+mantle}/M_{\oplus}$ posterior & $4.0^{+0.26}_{-0.27}$ & $23.41^{+1.12}_{-1.48}$ \\
$Z_{\rm env}$ prior & $\mathcal{U}(0.0,1.0)$ & $\mathcal{U}(0.0,1.0)$ \\
$Z_{\rm env}$ posterior & $0.30^{+0.14}_{-0.17}$ & $0.22^{+0.15}_{-0.14}$ \\
\end{tabular}
\tablefoot{Prior has been lowered on $M_{\rm core+mantle}$ for TOI-442\,b so that it was 90\% of the total mass, as described in text above. $\mathcal{U}(a,b)$ refers to a uniform distribution, ${\rm log\,}\mathcal{U}(a,b)$ to a log-uniform distribution and $\mathcal{N}(\mu,\sigma)$ to a normal distribution.}
\end{table}

\begin{figure*}
\centering
\includegraphics[width=1.0\hsize]{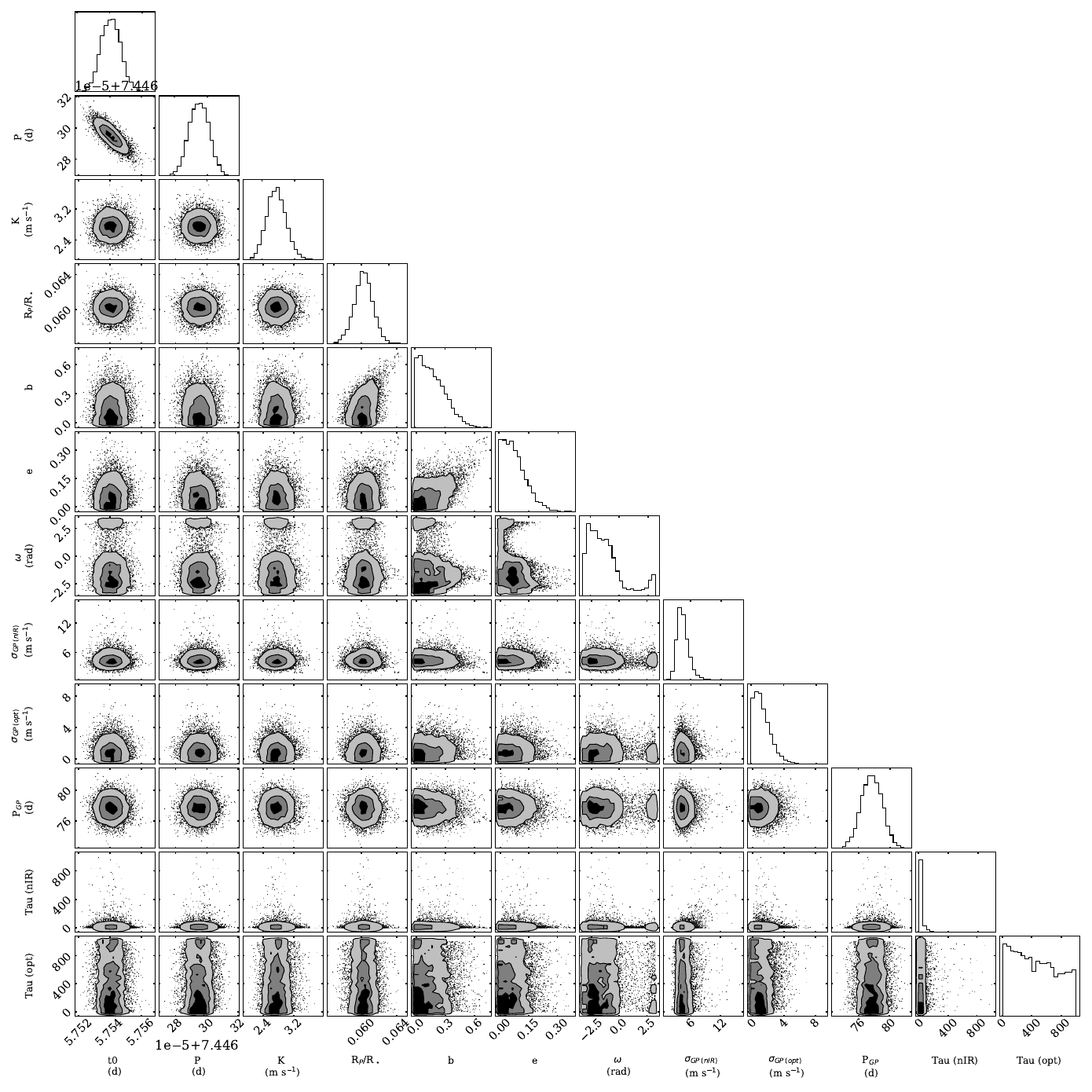}
\caption{Results from the posterior distribution sampling of TOI-4438, the shape of the distributions, and the correlations between parameters. Only the planetary and GP parameters are included. The contour levels show the 16th, 50th, and 84th quantiles.}
\label{fig:corner_4438}
\end{figure*}

\begin{figure*}
\centering
\includegraphics[width=1.0\hsize]{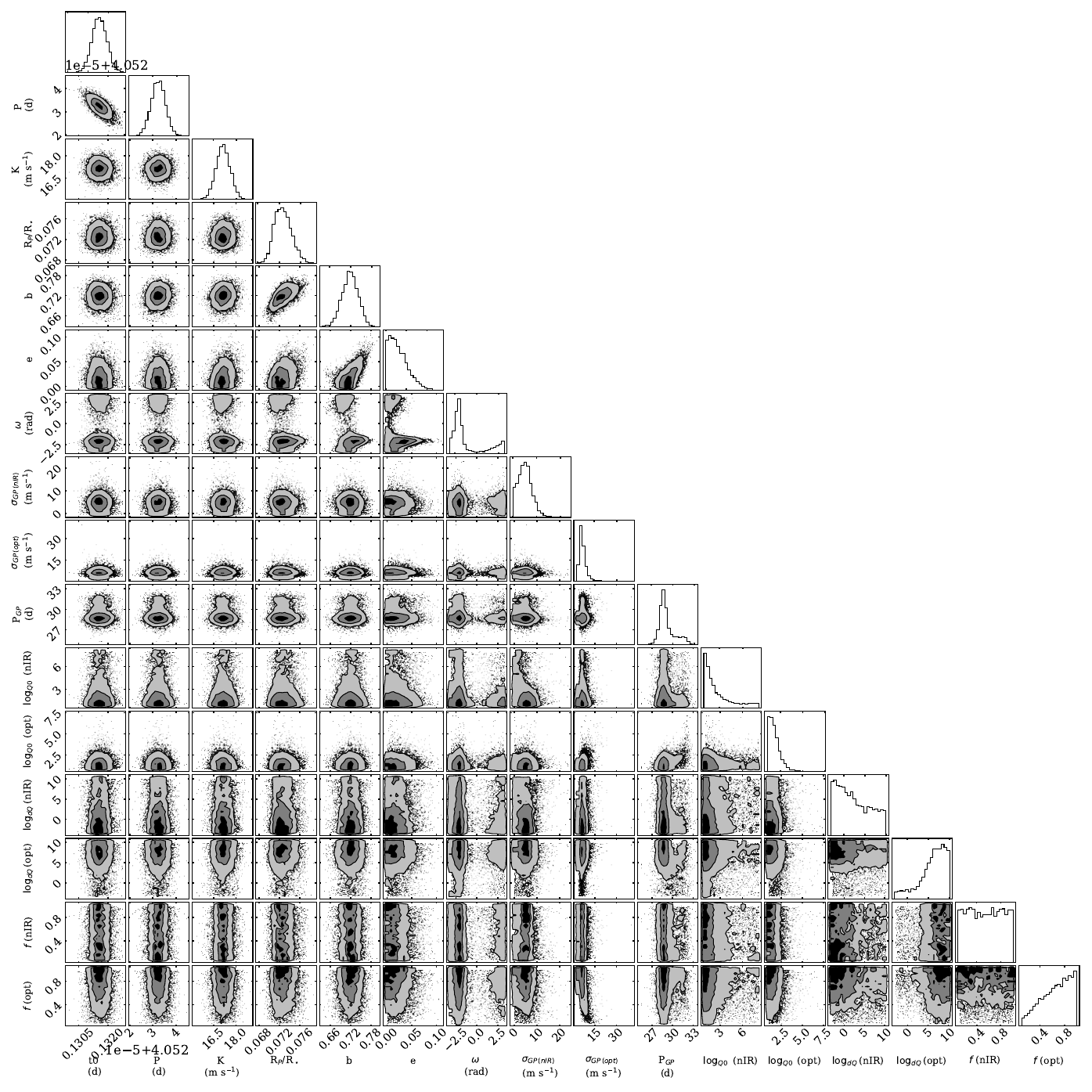}
\caption{Results from the posterior distribution sampling of TOI-442, the shape of the distributions, and the correlations between parameters. Only the planetary and GP parameters are included. The contour levels show the 16th, 50th, and 84th quantiles.}
\label{fig:corner_442}
\end{figure*}

\begin{figure*}
\centering
\includegraphics[width=0.8\hsize]{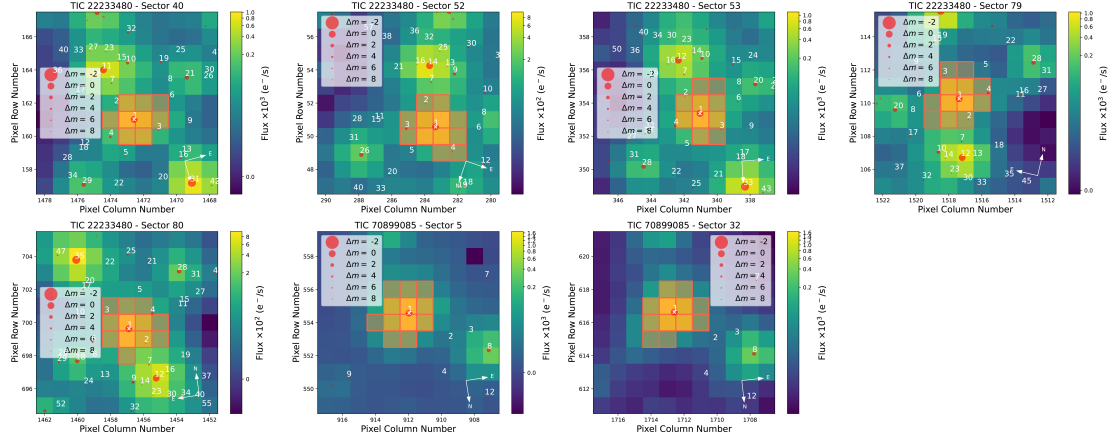}
\caption{TESS target pixel file images of all sectors used in our analysis. From left to right and top to bottom, the first five images correspond to sectors 40, 52, 53, 79, and 80 for TOI-4438, and the last two to sectors 5 and 32 for TOI-442. Pixels colored in red were used for the simple aperture photometry. The target star and nearby sources down to 8 G-band magnitudes fainter than the target stars are marked with numbered red circles at their Gaia DR3 positions \citep{Gaia2023}; number 1 is the target star. The size of the circles codes their relative magnitude with respect to the target star. The plots were made using the \texttt{tpfplotter} tool \citep{Aller2020}.}
\label{fig:tpf}
\end{figure*}

\begin{figure*}
\centering
   \includegraphics[width=0.9\hsize]{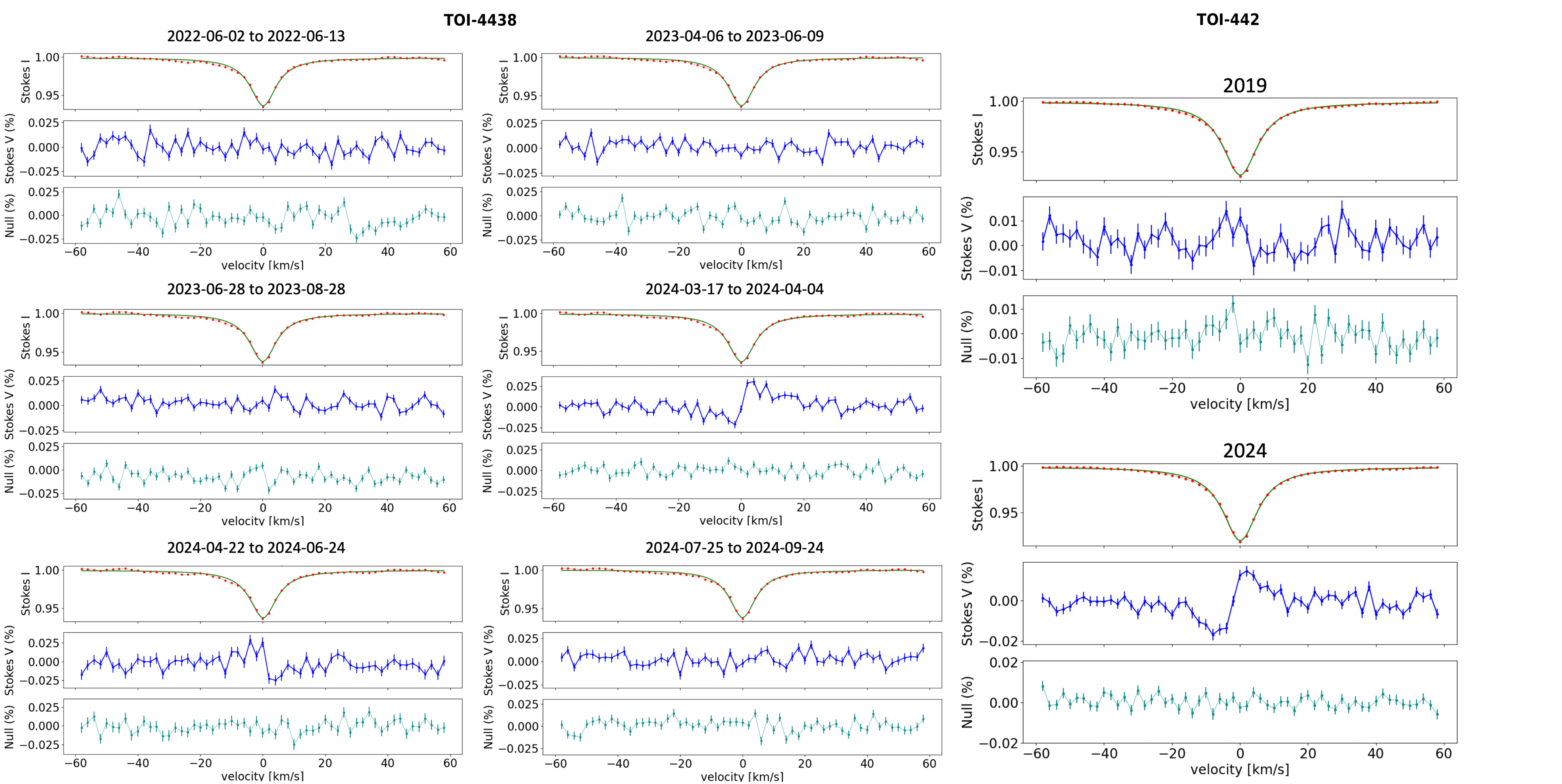}
    \caption{Median LSD profiles from SPIRou spectropolarimetric observations, showing Stokes~I with a Voigt-profile fit (green line), Stokes~V, and the null polarization profile. For TOI-4438 the median profiles are shown separately for the six different intervals, as indicated in the titles of each panel. For TOI-442 the median profiles are shown separately for the 2019 and 2024 observing seasons.}
    \label{fig:lsd_img1}
\end{figure*}

\section{Interior modeling details}
\label{app:interior}

The interior model previously presented and updated in \citet{Dorn2015, Dorn2017,luo_majority_2024} is built on the following assumptions. For the solid phase of the iron core, we used the equation of state (EOS) of hexagonal close packed iron \citep{hakim_new_2018, miozzi_new_2020}. For the liquid iron phase, we used the EOS from \citet{luo_majority_2024}. The silicate mantle is composed of three major species MgO, SiO$_2$, and FeO. We modeled the solid phase of the mantle using the predefined minerals and their equation of state in \citet{sotin2007mass}. The liquid mantle was modeled as a mixture of Mg$_2$SiO$_4$, SiO$_2$ and FeO \citep{melosh_hydrocode_2007,faik_equation_2018,ichikawa_ab_2020,stewart_shock_2020}. In all cases, the EOS of the different components were mixed using the additive volume law for ideal gases. Both the iron core and the silicate mantle were modeled as adiabatic. For this application, we fix an Earth-like core-to-mantle ratio of 0.325:0.675. This is assumed as degeneracy for interior parameters is generally high and the densitiy of sub-Neptunes is dominated by properties of the atmosphere layer.

The H$_2$-He-H$_2$O atmosphere layer was modeled using the analytic description by \citet{guillot_radiative_2010} and \citet{2014_Jin_planetarypopulation}, and consists of an irradiated layer on top of a non-irradiated layer in radiative-convective equilibrium. The planet intrinsic luminosity is calculated following \citep{mordasini_planetary_2020} and is a function of planet mass, atmospheric mass fraction ($M_{\rm atm}/M_{\rm tot}$) and system age. The water mass fraction is given by $Z_{\rm env}$, and the H/He ratio is set to solar. The two components of the atmosphere, H$_2$/He and H$_2$O, are mixed following the additive volume law. We used the EOS by \citet{1995_Saumon_EOS} for H$_2$/He and the ANalytic Equations of States (ANEOS, cf.  \citealt{1990_thompson_aneos}) for H$_2$O. The transit radius is evaluated where the chord optical depth is $\tau_\text{ch}=0.56$ \citep{2008Lecavelier}.

For the inference, we used Polynomial Chaos Kriging surrogate modeling to approximate the global behavior of the forward model and replace it within the MCMC framework \citep{DeWringer2026}. In all cases, surrogate errors are well below observational uncertainties and are propagated in the likelihood, including model uncertainty.

\section{SPIRou spectropolarimetry}
\label{apx:spirou_data}

The spectroscopic and spectropolarimetric time series obtained with SPIRou for TOI-442 and TOI-4438 are available electronically at the CDS. 

The full tables contain the following information. Column~1 lists the barycentric time of observation expressed as reduced Julian date (RJD). Column~2 gives the measured radial velocity (RV), and Column~3 its associated uncertainty. Columns~4 and~5 provide the differential line width (dLW) and its uncertainty. Columns~6 and~7 give the differential temperature indicator (dTEMP) and its uncertainty. Columns~8 and~9 list the longitudinal magnetic field $B_\ell$ derived from the spectropolarimetric observations and its uncertainty.

\end{document}